\documentclass[%
reprint,
superscriptaddress,
showkeys,
amsmath,amssymb,
aip,apl,%
]{revtex4-1}
\pdfoutput=1
\usepackage{placeins}
\usepackage{graphicx}
\usepackage{comment}
\usepackage{gensymb}
\usepackage{amsmath}
\usepackage[caption=false]{subfig}
\usepackage{framed}
\usepackage[normalem]{ulem}
\usepackage[export]{adjustbox}
\usepackage{dcolumn}
\usepackage{bm}
\usepackage{array,ragged2e}
\newcolumntype{P}[1]{>{\Centering\hspace{0pt}}p{#1}}
\usepackage{subfig}
\newsavebox{\measurebox}
\usepackage[utf8]{inputenc}
\usepackage[T1]{fontenc}
\usepackage{mathptmx}
\usepackage[dvipsnames]{xcolor}

\begin{document}
\preprint{APS/123-QED}
\title[Suppression of the Shear Raman Mode in Defective Bilayer MoS$_2$]{Suppression of the Shear Raman Mode in Defective Bilayer MoS$_2$}
\author{Pierce~Maguire}
\affiliation{School of Physics, Trinity College Dublin, Dublin 2, Ireland, D02 PN40}
\affiliation{AMBER Centre, CRANN Institute, Trinity College Dublin, Dublin 2, Ireland, D02 PN40}

\author{Clive~Downing}
\affiliation{AMBER Centre, CRANN Institute, Trinity College Dublin, Dublin 2, Ireland, D02 PN40}
\affiliation{School of Chemistry, Trinity College Dublin, Dublin 2, Ireland, D02 PN40}

\author{Jakub~Jadwiszczak}
\affiliation{School of Physics, Trinity College Dublin, Dublin 2, Ireland, D02 PN40}
\affiliation{AMBER Centre, CRANN Institute, Trinity College Dublin, Dublin 2, Ireland, D02 PN40}

\author{Maria~O'Brien}
\affiliation{AMBER Centre, CRANN Institute, Trinity College Dublin, Dublin 2, Ireland, D02 PN40}
\affiliation{School of Chemistry, Trinity College Dublin, Dublin 2, Ireland, D02 PN40}

\author{Darragh~Keane}
\affiliation{AMBER Centre, CRANN Institute, Trinity College Dublin, Dublin 2, Ireland, D02 PN40}
\affiliation{School of Chemistry, Trinity College Dublin, Dublin 2, Ireland, D02 PN40}

\author{John~B.~McManus}
\affiliation{AMBER Centre, CRANN Institute, Trinity College Dublin, Dublin 2, Ireland, D02 PN40}
\affiliation{School of Chemistry, Trinity College Dublin, Dublin 2, Ireland, D02 PN40}

\author{Georg~S.~Duesberg}
\affiliation{AMBER Centre, CRANN Institute, Trinity College Dublin, Dublin 2, Ireland, D02 PN40}
\affiliation{School of Chemistry, Trinity College Dublin, Dublin 2, Ireland, D02 PN40}
\affiliation{Institute of Physics, EIT 2, Faculty of Electrical Engineering and Information Technology, Universit{\"a}t der Bundeswehr M{\"u}nchen, Werner-Heisenberg-Weg 39, 85577 Neubiberg, Germany}

\author{Valeria~Nicolosi}
\affiliation{AMBER Centre, CRANN Institute, Trinity College Dublin, Dublin 2, Ireland, D02 PN40}
\affiliation{School of Chemistry, Trinity College Dublin, Dublin 2, Ireland, D02 PN40}

\author{Niall~McEvoy}
\affiliation{AMBER Centre, CRANN Institute, Trinity College Dublin, Dublin 2, Ireland, D02 PN40}
\affiliation{School of Chemistry, Trinity College Dublin, Dublin 2, Ireland, D02 PN40}

\author{Hongzhou~Zhang}
\email{Hongzhou.Zhang@tcd.ie}
\affiliation{School of Physics, Trinity College Dublin, Dublin 2, Ireland, D02 PN40}
\affiliation{AMBER Centre, CRANN Institute, Trinity College Dublin, Dublin 2, Ireland, D02 PN40}
\date{\today}
\begin{abstract}
We investigate the effects of lattice disorders on the low frequency Raman spectra of bilayer MoS$_2$. The bilayer MoS$_2$ was subjected to defect engineering by irradiation with a 30 keV He$^+$ ion beam and the induced morphology change was characterized by transmission electron microscopy. With increasing ion dose the shear mode is observed to red-shift and it is also suppressed sharply compared to other Raman peaks. We use the linear chain model to describe the changes to the Raman spectra. Our observations suggest that crystallite size and orientation are the dominant factors behind the changes to the Raman spectra.
\end{abstract}
\maketitle
As research on transition metal dichalcogenides (TMDs) blooms \cite{Novoselov2016,Jariwala2016,Kang2017,Sung2017}, it is crucial to understand the interfacial dynamics of these layered semiconductors in the few-layer limit. Low-frequency interlayer vibrational Raman modes serve as a fingerprint of the interlayer interactions between van der Waals-bonded sheets in both homo- and heterostructures \cite{Boukhicha2013, Lui2015, Zhang2016a, Li2016}. The low-frequency Raman modes of MoS$_2$ include the in-plane shear mode (SM) and the out-of-plane layer breathing mode (LBM). Layer thickness, stacking order and angular lattice mismatch can be directly probed in vertically-assembled heterostructures of TMDs by investigating changes to these modes \cite{Zhao2013,Zhang2013,Lui2015,Zhang2015,Lin2016a,Song2016}. On the other hand, Raman spectroscopy has remained a high-throughput and non-destructive way of analyzing defects in MoS$_2$ flakes in the last decade. In particular, the $E^{1}_{2g}$ ($\sim$385 cm$^{-1}$) and $A_{1g}$ ($\sim$403 cm$^{-1}$) modes are routinely used to identify layer thickness in exfoliated and deposited MoS$_2$ samples \cite{Lee2010,Li2012,Zhang2013,OBrien2015,Zeng2012}. Relative changes in intensities, positions or widths are used to characterise the lattice modification of MoS$_2$ \cite{Fox2015,Ko2016,Choudhary2016,Maguire2018}.
However, the effects of lattice defects on the low-frequency ($<$50 cm$^{-1}$) modes of few-layer MoS$_2$ have not been explored, even while low-frequency Raman spectroscopy should play a vital role in the characterization of interfacial dynamics of defective layered semiconductors. 

Herein we analyze the low-frequency modes of the Raman spectrum of bilayer MoS$_2$ as a function of increasing disorder introduced by He$^+$ irradiation at a beam energy of 30 keV. The helium ion microscope has been established as a tool for site-specific modification of 2D materials \cite{Fox2015, Zhou2016a, Nanda2015, Iberi2016, Stanford2016, Stanford2017b, Nanda2017a}. It has been used to precisely introduce defects \cite{Maguire2018}, fabricate nanoribbons and periodic arrays of nanodots \cite{Fox2015,Zhou2016a} and introduce doping on the nanoscale \cite{Nanda2015}. 

MoS$_2$ was prepared using a previously reported chemical vapor deposition (CVD) technique \cite{OBrien2014}. Bilayer flakes of MoS$_2$ were identified on the SiO$_2$ surface by optical contrast and Raman spectroscopy \cite{Lee2010, Gomez2012}. The \textit{Zeiss ORION NanoFab} microscope was used to irradiate arrays of 4$\times$4 $\mu$m$^2$ regions in MoS$_2$ with He$^+$ at an energy of 30 keV and an angle of incidence of 0$^{\circ}$. These regions received doses ranging from 8$\times$10$^{13}$ to 2$\times$10$^{16}$ He$^+$ cm$^{-2}$. Further details of the ion irradiation are provided in the supplementary material. Ex-situ Raman and PL spectroscopy was carried out using a \textit{WITec Alpha 300R} system (532 nm laser) with a 1800 lines/mm diffraction grating and a 100$\times$ objective (NA=0.95) (spot size $\sim$0.3 $\mu$m). The instrument was equipped with a rayshield coupler to detect Raman signal close to the Rayleigh line. The laser power was $\sim$170 $\mu$W or less to minimise damage to the samples. Raman and PL maps were generated by taking four spectra per $\mu$m in both the x and y directions over large areas. The acquisition time was 0.113 s. The spectra from a desired region were acquired by averaging. Raman, PL and optical images are shown in Figure S1. Lorentzian distributions were fitted to the Raman peaks as demonstrated in Figure S4.

Figure 1 shows representative Raman spectra. In the low frequency region (Fig 1(a)), the SM and LBM peaks are centred at $\sim$41.0 cm$^{-1}$ and $\sim$23 cm$^{-1}$ respectively, which are in excellent agreement with the reported values for bilayer MoS$_2$ \cite{Zhao2013,Zeng2012,Liang2017b}. For increasing dose, the SM is clearly observed to red-shift and broaden while the LBM does not appear to change. At higher doses ($\sim$1$\times 10^{16}$ He$^+$ cm$^{-2}$) the low frequency modes become indistinguishable from noise. In the high frequency region (Fig 1b), the characteristic $E_{g}$ and $A_{1g}$ peaks of bilayer MoS$_2$ are found at $384$  cm$^{-1}$ and $406.6$ cm$^{-1}$ respectively, thus exhibiting a high frequency peak separation of $\sim$22.6 cm$^{-1}$ typical of CVD bilayer MoS$_2$ \cite{Lee2010, Liu2014}. With increasing ion dose the the $A_{1g}$ peak is observed to broaden while the $E_{g}$ peak broadens, splits and is shifted downward in energy as has been reported before in similar conditions \cite{Mignuzzi2015, Klein2017, Maguire2018}.
\begin{figure}
\centering
\subfloat[]{\label{HeSpectraLow}\includegraphics[height=1.65in]{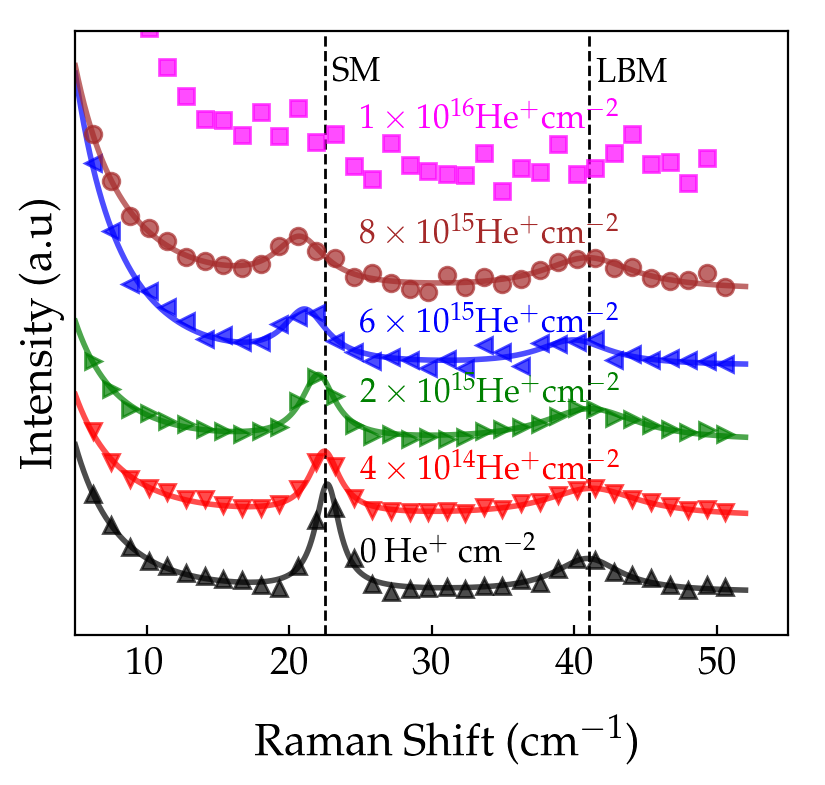}}
\subfloat[]{\label{HeSpectraHigh}\includegraphics[height=1.65in]{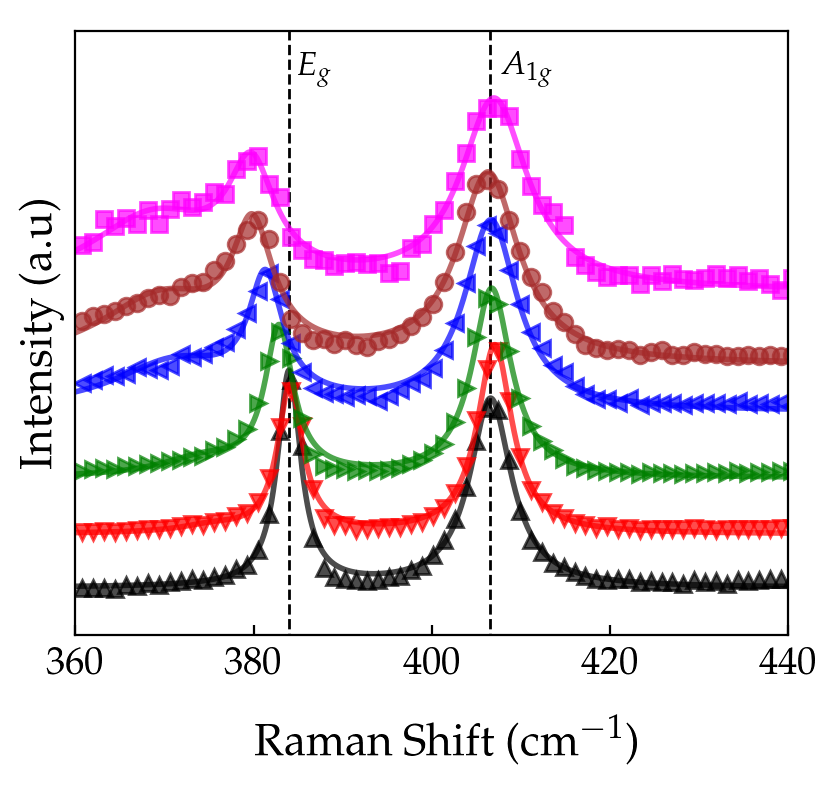}}
\caption{A representative selection of Raman spectra of bilayer MoS$_2$ excited by a 532 nm laser after irradiation with He$^+$ at 30 keV with a 0$^{\circ}$ angle of incidence. The evolution of the spectra with increased ion dose is shown ascending from the bottom in black to the top in magenta. (a) shows the low frequency region in which the SM and LBM peaks are labelled. (b) shows the high frequency region in which the $E_{g}$ and $A_{1g}$ peaks are labelled. In both high and low frequency ranges each spectrum is normalized to its $A_{1g}$ peak. The intensity is rescaled between high and low frequency spectra for maximum visibility.}
\label{fig:ramanspectra}
\end{figure}

Figure 2 shows photoluminescence (PL) spectra acquired from the same regions of bilayer MoS$_2$. We observe peaks at $\sim$1.85 eV and $\sim$2.01 eV, corresponding to the $A$ and $B$ direct excitonic transitions respectively \cite{Splendiani2010}. The increased disorder is strongly associated with a lower quantum efficiency. The decrease in the $A$ exciton peak intensity is accompanied by a blue-shift which is attributed to strain caused by ion-induced defects \cite{Scheuschner2014}. Above a dose of $\sim$2$\times 10^{15}$ He$^+$ cm$^{-2}$ we can no longer observe a clearly defined emission peak. 
\begin{figure}
\centering
\subfloat{\label{sub:HeSpectraPl}\includegraphics[height=1.85in]{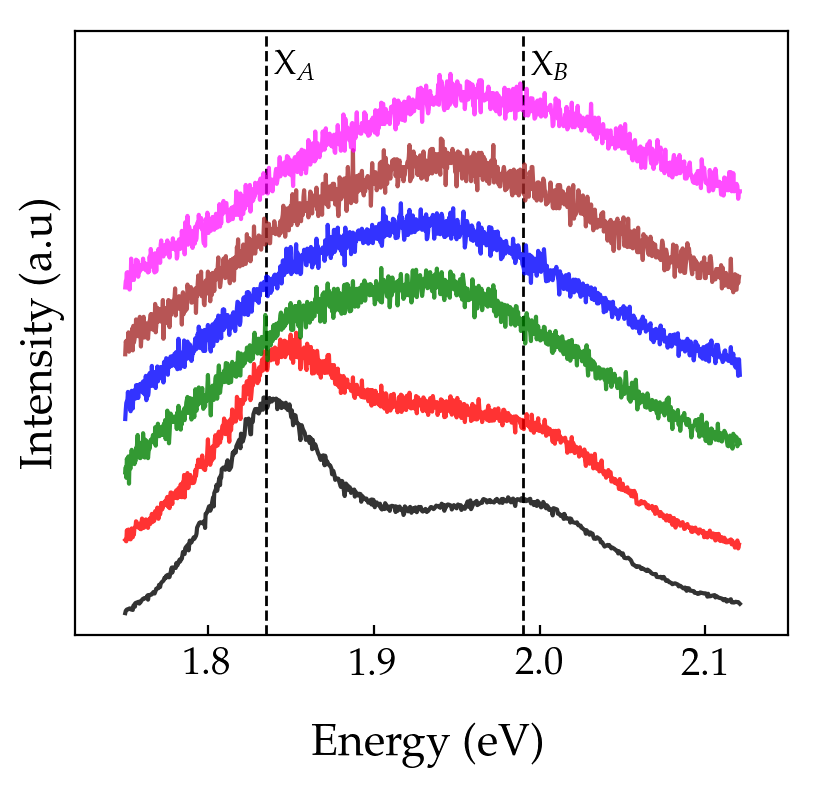}}
\caption{PL spectra of the same regions with the same color-dose correspondence as in Figure 1. $X_A$ and $X_B$ label the A and B excitons respectively. The A exciton emission is observed to decline rapidly with increasing dose. The spectra are normalized to the maximum of each to allow comparison of the two peaks. The spectra are presented without normalisation in Figure S2.} 
\label{fig:plspectra}
\end{figure}

Figures 3(a) and (b) show the position and width respectively of the low frequency Raman peaks as a function of ion dose (analysis of the high frequency peaks is presented in Figure S3). The SM clearly shifts downward in position while it broadens. However, the LBM does not appear to change systematically in position or width. The measured SM frequency, $\omega$, is given by
\begin{equation}
\omega = \omega_0 - 2.04\times10^{-8}\sqrt{S} 
\label{eq:omega}
\end{equation}
where $\omega_0=22.8$ cm$^{-1}$ is the frequency of the non-irradiated bilayer MoS$_2$ and $S$ is the ion dose (He$^+$ cm$^{-2}$). The measured width, $\Gamma$, is given by
\begin{equation}
\Gamma = \Gamma_0 + 0.247\sqrt{S}
\label{eq:gamma}
\end{equation}
where $\Gamma_0 =1.9$ cm$^{-1}$ is the width of the undamaged bilayer MoS$_2$.
\begin{figure*}
\centering
\subfloat[]{\label{sub:LPosition}\includegraphics[height=2.10in]{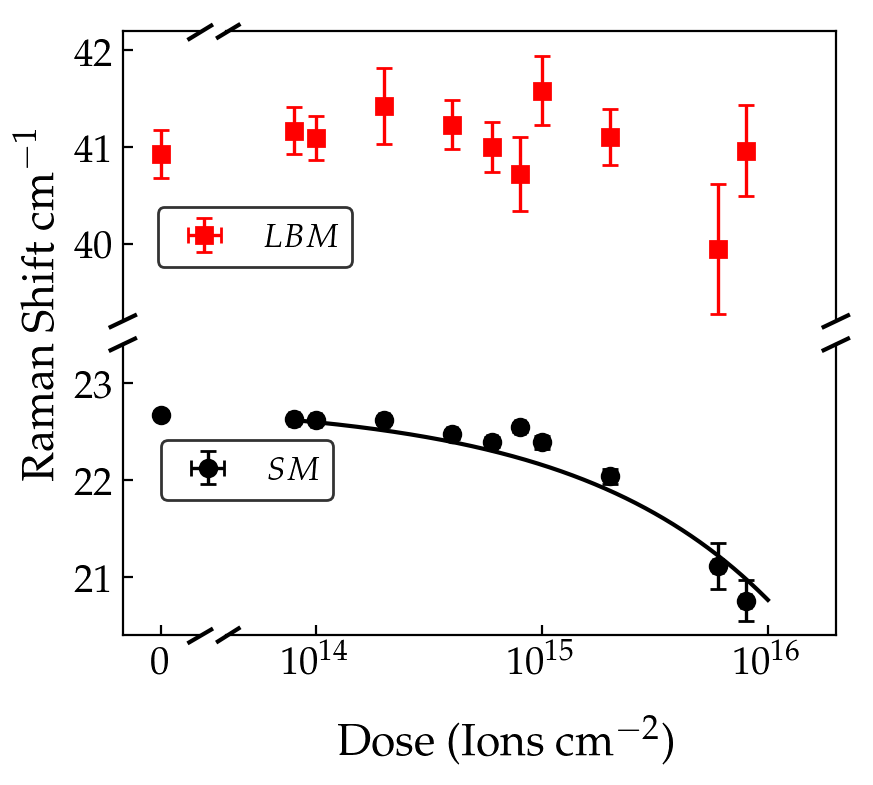}}
\subfloat[]{\label{sub:LWidth}\includegraphics[height=2.10in]{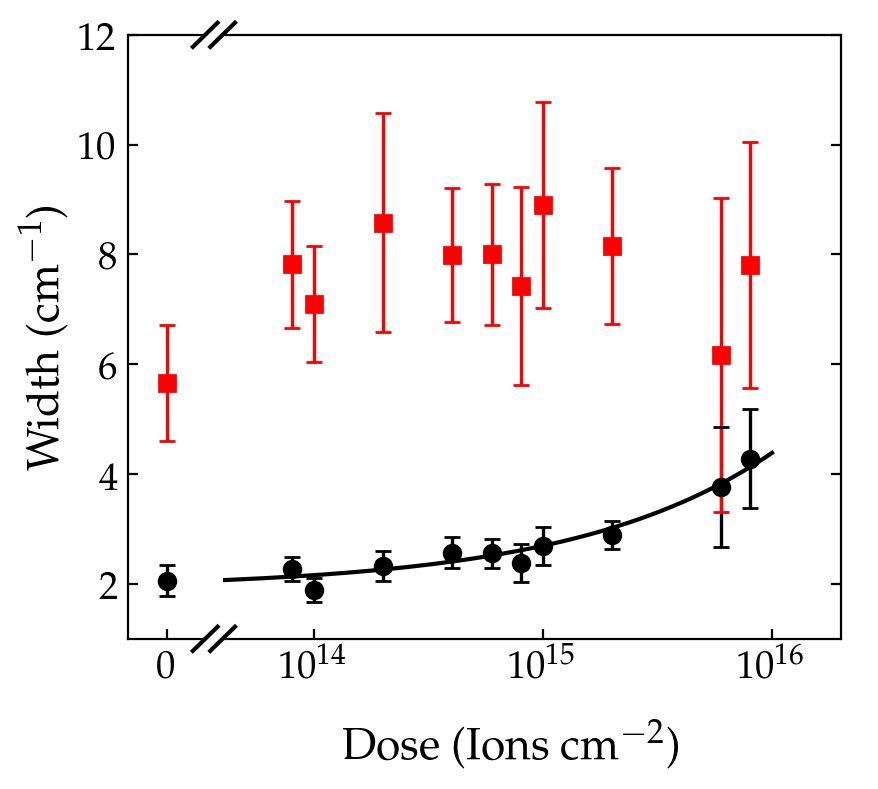}}
\subfloat[]{\label{sub:LIntRatio}\includegraphics[height=2.10in]{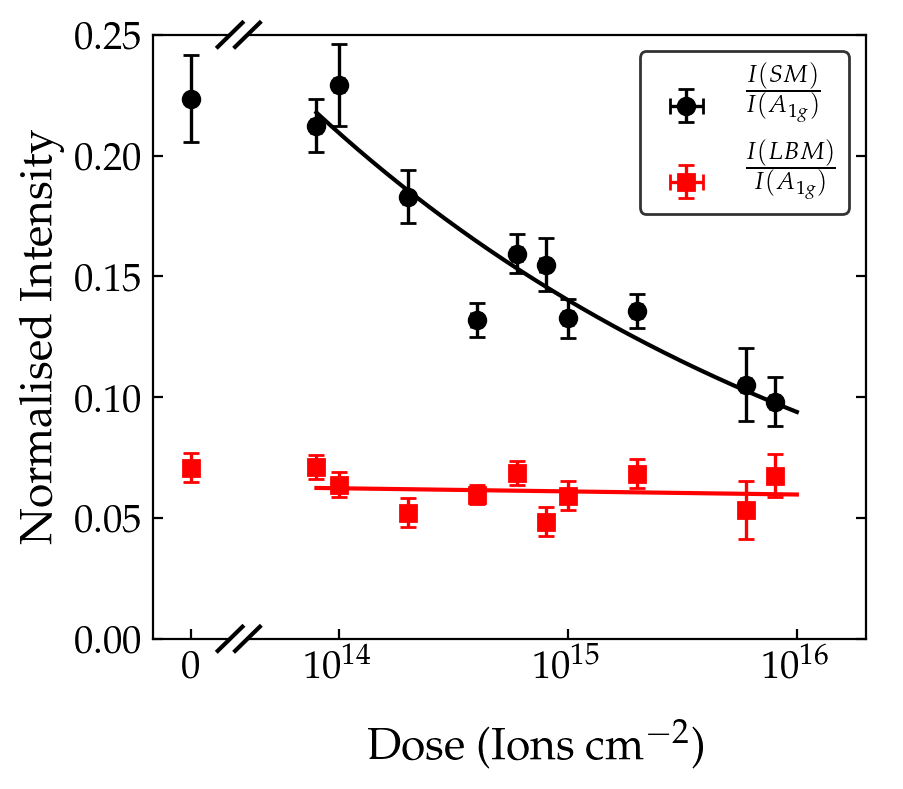}}\quad
\caption{Evolution of the low frequency Raman modes as a function of ion dose. (Error bars are from the fitting error.) 
(a) Shows the change in frequency of the two low frequency modes.
(b) Shows the change in width of the two low frequency modes.
(c) Shows the ratio of the intensity of the two low frequency modes to the $A_{1g}$ peak.}
\label{fig:Ld}
\end{figure*}

Figure 3(c) shows the heights of the low frequency modes, $I(X)$ where $X$ = SM or LBM, normalized to the respective $A_{1g}$ intensity, I($A_{1g}$). While the intensity of $I(LBM)/I(A_{1g})$ is virtually constant, $I(SM)/I(A_{1g})$ is observed to diminish with increasing dose. The SM is clearly suppressed more rapidly than other peaks by increasing disorder. The SM intensity ratio varies with the dose $I(SM)\propto{I(A_{1g})/S^d}$, where $d = 0.17$ is a fitting parameter.

Figure 4(a) is a high resolution transmission electron microscopy (HRTEM) image of mechanically exfoliated bilayer MoS$_2$ showing clearly defined layers with uniform interlayer spacing (see supplementary material including Figure S5 for TEM sample preparation). Above and below the MoS$_2$, the protective platinum layer and the SiO$_2$ substrate are visible, respectively. Figure 4(b) shows a region of the same sample which has been irradiated with 6$\times 10^{15}$ He$^+$ cm$^{-2}$ which is in the middle-to-high part of the range of doses used in our spectoscopic experiments. An additional amorphous hydrocarbon layer of helium ion beam-induced deposition (IBID) is visible on top of the MoS$_2$. A clear increase in the surface asperity is noted after irradiation. In certain regions, MoS$_2$ is still visibly present but separate layers cannot be resolved. This may suggest amorphization or local twisting of the layers such that they are imaged off-axis.

Figure 4(c) is a histogram of interlayer separation evaluated from the two TEM images, using a fitting process described in the supplementary material. There appears to be very little change in the mean layer separation after irradiation. However, the disordered distribution is clearly more diffuse and skewed towards higher values, reflected in the increased kurtosis and skewness respectively.

\begin{figure}
\centering
\sbox{\measurebox}{%
  \begin{minipage}[b][\ht\measurebox][s]{.25\textwidth}
  \subfloat
    []
   {\label{sub:04notirradiatedarea}\includegraphics[height=1.65cm]{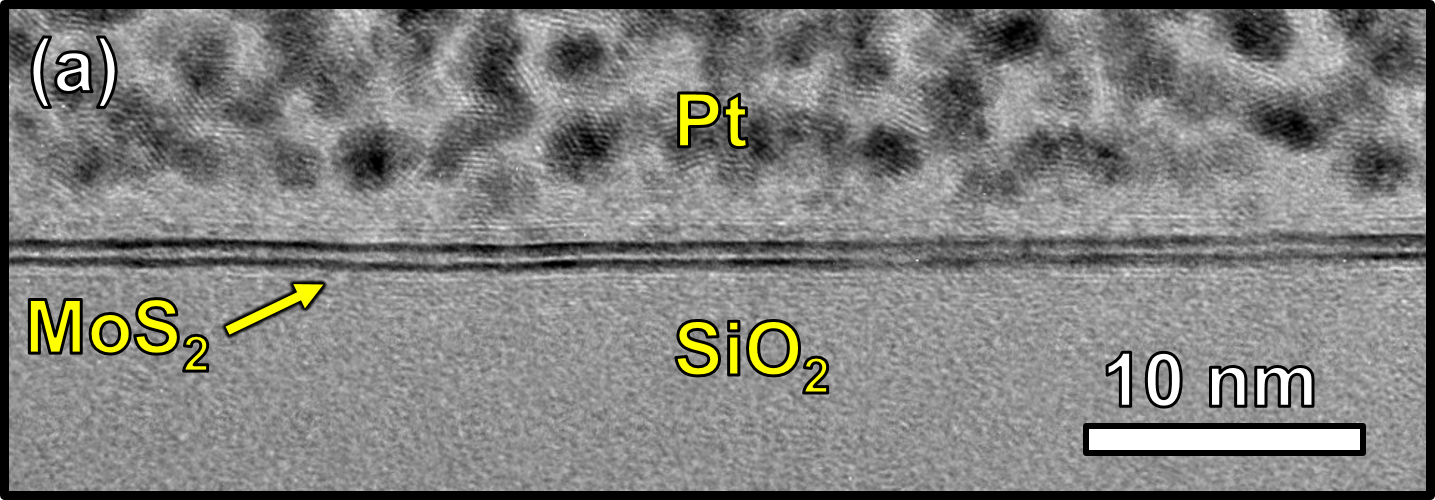}} 
    \vfill 
   \subfloat
  []  {\label{sub:02irradiatedarea}\includegraphics[height=1.65cm]{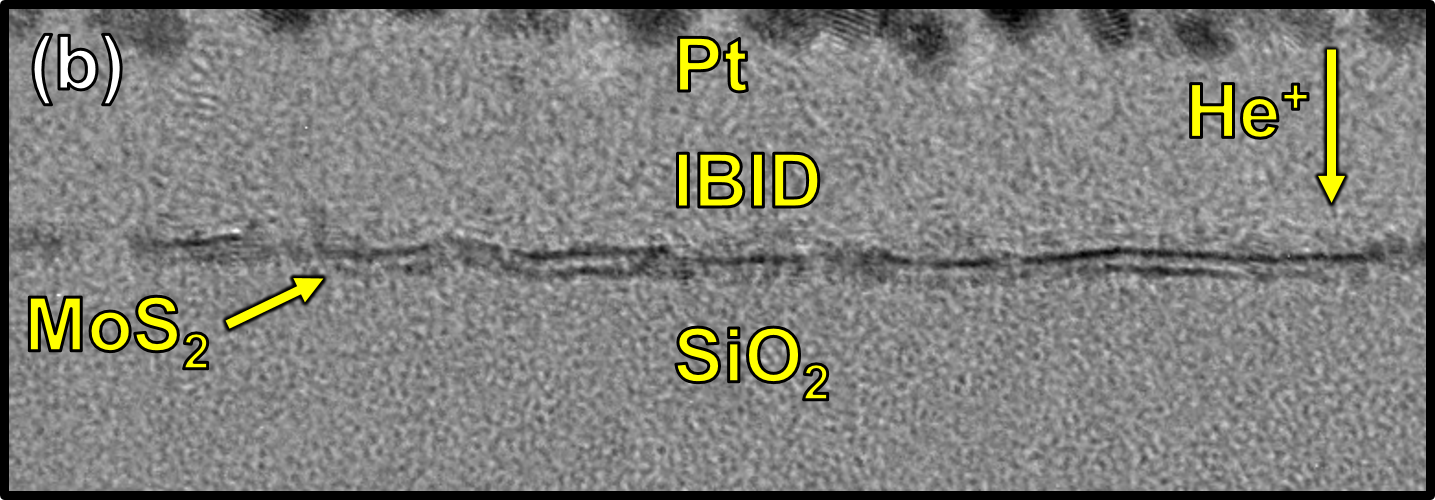}}
\end{minipage}}
\usebox{\measurebox}\quad
\begin{minipage}[b][\ht\measurebox][s]{.15\textwidth}
\centering
\subfloat
  []
  {\label{sub:histpristine}\includegraphics[height=4.15cm]{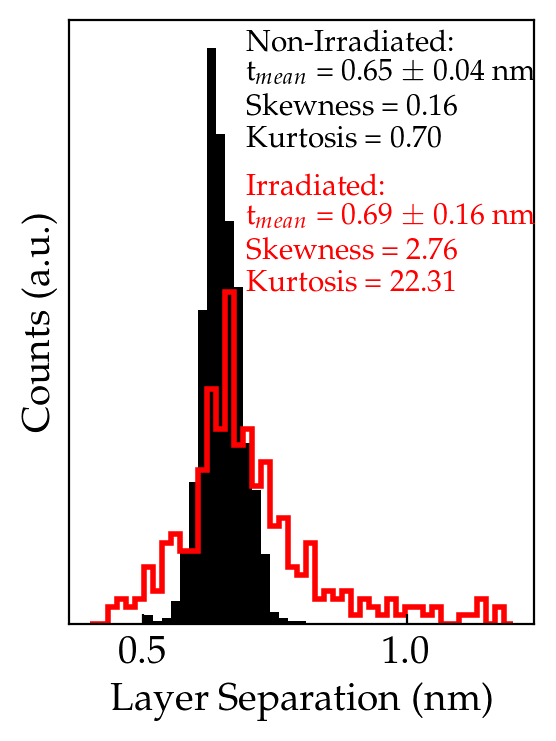}}
\end{minipage}
\vspace{4.15 cm}
\caption{Cross-sectional HRTEM images and analysis of bilayer MoS$_2$ which have the same scale. (a) and (b) are images of a non-irradiated region and a region irradiated with 6$\times 10^{15}$ He$^+$ cm$^{-2}$ respectively. Both (a) and (b) have the same scale. The labels and arrows indicate the various materials and the direction of the ion beam irradiation. A histogram of the interlayer separation for non-irradiated and irradiated MoS$_2$. The filled black bars represent the layer separation of the non-irradiated region (from Figure 4(a)) and the unfilled red bars represent the layer separation of the irradiated region (from Figure 4(b). The mean, skewness and kurtosis of each distribution are presented in the respective colors.
}
\label{fig:Test}
\end{figure}

A linear chain model (LCM) has been used to describe interlayer restoring forces by treating each layer as a rigid ball \cite{Zeng2012,Tan2012}. The model has accurately predicted the frequencies of the low frequency modes in pristine MoS$_2$ as a function of layer number.
The single layer mass per unit area ($\mu_0$) of pristine MoS$_2$ is 3.03$\times 10^{-26}$ kg \AA$^{-2}$.
The effects of irradiation-induced sputtering are accounted for in $\mu$ by applying an approximate correction related to the ion dose. Given the dose, $S$, and the sputter yield, $\gamma$ (=0.007 from previous reports\cite{Maguire2018, Kretschmer2018,Parkin2016}), then $\mu (S)$ for irradiated MoS$_2$ is given by $\mu (S) = \mu_0 (1 - \gamma S/n)$ where n is the atomic density. This gives the sputtering-corrected LCM as:
\begin{equation}
\omega (S) = \frac{1}{\sqrt[]{2}\pi c} \sqrt[]{\frac{\alpha}{\mu_0 (1 - \frac{\gamma S}{n})}} \sqrt[]{1 + \cos(\frac{\pi}{N})}
\label{eq:massadjustedLCM}
\end{equation}
where $c$ is the speed of light in cm s$^{-1}$, $N$ is the layer number (fixed at 2 in this work), $\alpha$ is the interlayer force constant per unit area which is related to the interlayer separation, $t$, and the shear modulus, $C$ by $\alpha = C/t$. The separation of two non-irradiated layers in bilayer MoS$_2$ measured from our TEM results is $t=0.65\pm 0.04$ nm, in good agreement with literature \cite{Zeng2012}. The shear modulus has been reported as $C$ $\simeq$ 17.9 GPa \cite{Zeng2012}. For $S$ = 0, the shear mode position is calculated to be 23.2 cm$^{-1}$ which is consistent with our experimental value of 22.8 cm$^{-1}$.

The two key findings of our Raman spectroscopy experiments clear from Figures 1 and 2 are related to the shear mode, i.e. the drop in the normalized intensity ($\sim$2.3$\times$) and the frequency red-shift ($\Delta \omega \simeq{}1.8$ cm$^{-1}$) with ion irradiation. Small twisting angles can introduce a periodicity mismatch between layers, altering the stacking configuration and causing a sharp decay of SM intensity and a downward shift in position \cite{Huang2016,Liang2017a}. The twisting is evident in the TEM observation (Figure 4). However, the twisting alone cannot explain the observed peak shift (see Figure S10). 
Reductions in the shear modulus also contribute to a red-shift in the SM. This is due to the reduced crystallinity in MoS$_2$ under ion irradiation. The reduced crystallinity leads to better lubricating properties, attributed to a greater tendency to exfoliate and therefore a reduced shear modulus \cite{Lahouij2012}. To further corroborate this, we studied the effect of varying the pixel spacing on the SM with a new set of irradiation experiments. We used a fixed dose of 6$\times$10$^{15}$ cm$^{-2}$ but we varied the pixel spacing from 1 to 16 nm. The probe size was $\sim$4 nm (see Figure S7) \cite{Rueden2017}. The resulting Raman spectra are presented in Figure 6(a). Figure 6(b) shows that as the pixel spacing decreases, the red-shift of the SM is observed to increase (total effect is approximately 0.5 cm$^{-1}$). As pixel spacings decrease, the total undamaged area is reduced. This suggests that regions of defective MoS$_2$ may have a reduced shear modulus in addition to the twisting effects.

\begin{figure}
\centering
\subfloat[]{\label{sub:HeSpectraSize}\includegraphics[height=1.95in]{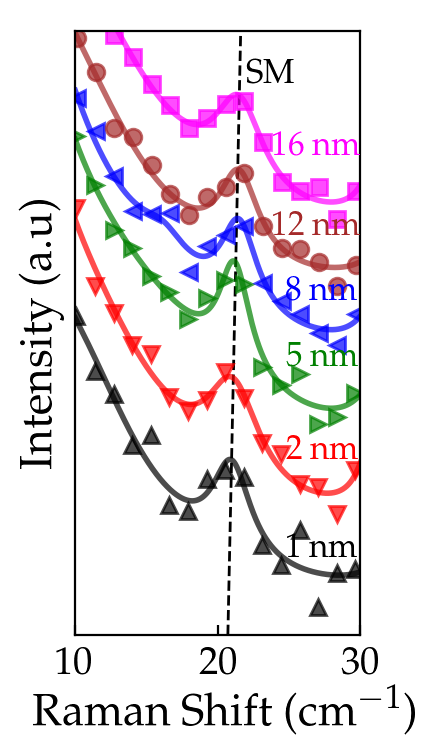}}
\subfloat[]{\label{sub:LPositionpix}\includegraphics[height=1.95in]{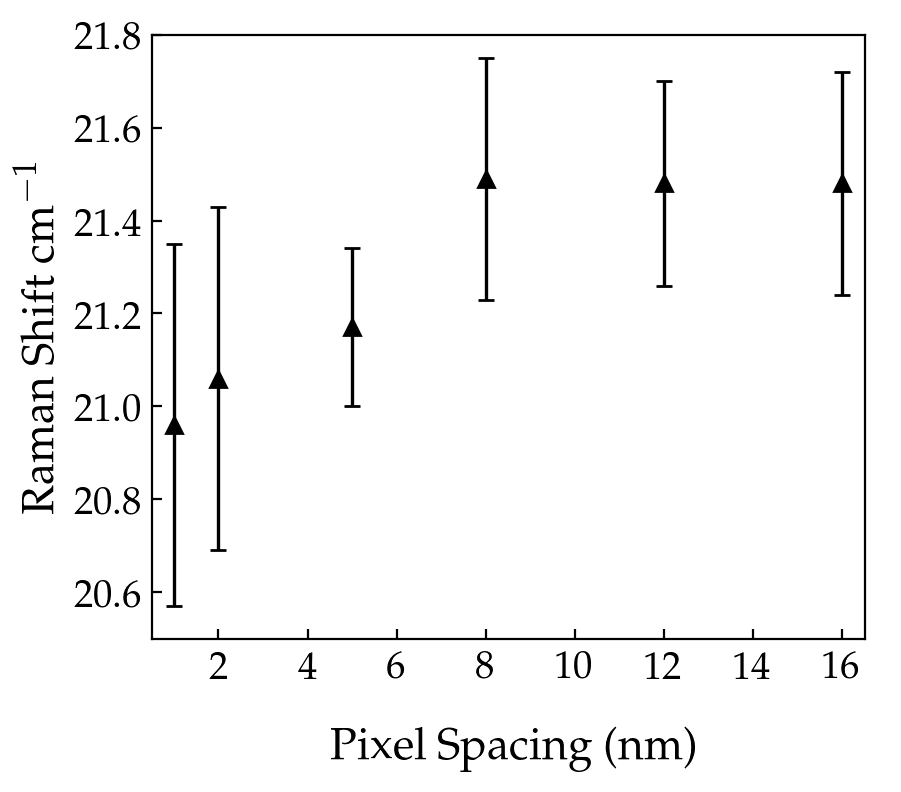}}
\caption{Evolution of the SM as a function of ion pixel spacing at a fixed dose of 6 $\times$ 10$^{15}$ cm$^{-2}$ and with a probe size of $\sim$4 nm. (a) shows the SM with the different pixel spacings labelled. 
(b) shows the change in position of the SM as a function of the irradiation pixel spacing.}
\label{fig:pixels}
\end{figure}

In conclusion, we explored the effects of increased disorder on the low and high frequency Raman peaks of bilayer MoS$_2$. 
In the low frequency range, we noticed that the shear mode red-shifted and dropped in intensity with increased irradiation dose. We used TEM imaging to investigate and we suggest that these changes can be attributed to a mixture of local twisting of layers and a decline in the shear modulus due to reduced crystallite sizes.

Supplementary material is available at **.
The authors thank the staff at the Advanced Microscopy Laboratory (AML), CRANN, Trinity College Dublin. We acknowledge support from the following grants: Science Foundation Ireland [grant numbers: 12/RC/2278,  15/SIRG/3329, 11/PI/1105, 07/SK/I1220a, 15/IA/3131, 08/CE/I1432 and GOIPG/2014/972]. 
\clearpage
\renewcommand{\theequation}{S\arabic{equation}}
\renewcommand{\thetable}{S\arabic{table}}
\renewcommand{\thefigure}{S\arabic{figure}}
\setcounter{figure}{0}
\setcounter{equation}{0}

\onecolumngrid
\begin{LARGE}
Supplementary Information
\end{LARGE}
\FloatBarrier
\section{Maps and PL}
\begin{figure*}
\centering
\subfloat[]{\label{sub:ramanmap}\includegraphics[height=2.40in]{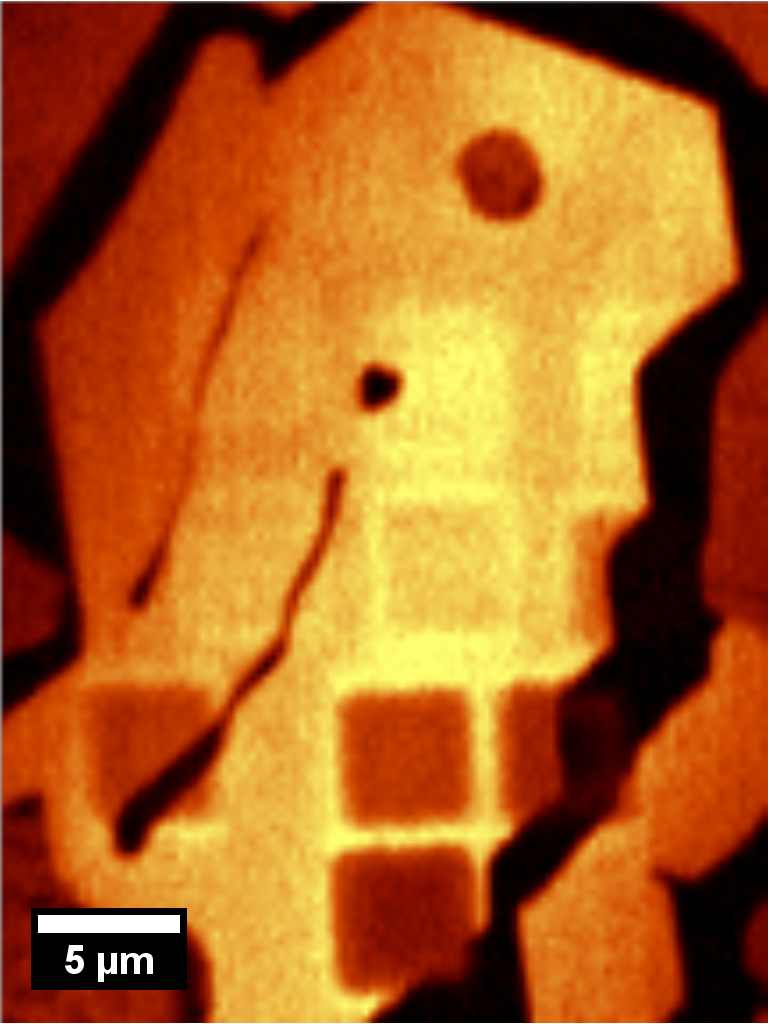}}\qquad
\subfloat[]{\label{sub:plmap}\includegraphics[height=2.40in]{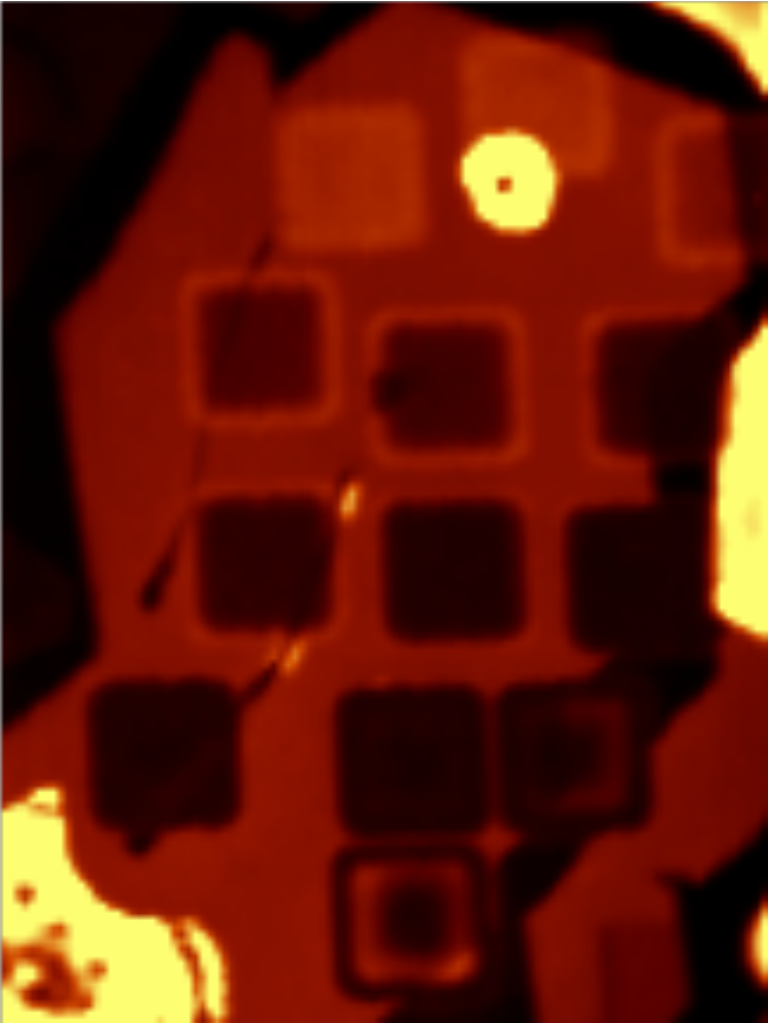}}\qquad
\subfloat[]{\label{sub:optical}\includegraphics[height=2.40in]{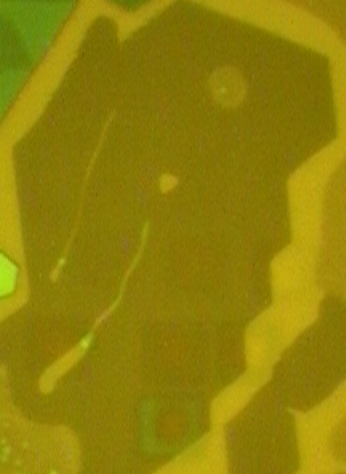}}
\caption{Images of He$^+$ irradiated bilayer MoS$_2$.
(a) is a Raman map of the intensity of the A$_{1g}$ peak at 406 cm$^{-1}$ with a spectral window of 4 cm$^{-1}$. The discoloured squares are regions irradiated with different doses.
(b) is a PL map of the A exciton centred at 1.8 eV with a spectral window of 0.1 eV. 
(c) shows an optical image of the same region.
}
\label{fig:maps}
\end{figure*}

\begin{figure*}
\centering
\subfloat{\label{sub:HeSpectraPLnonorm}\includegraphics[height=1.90in]{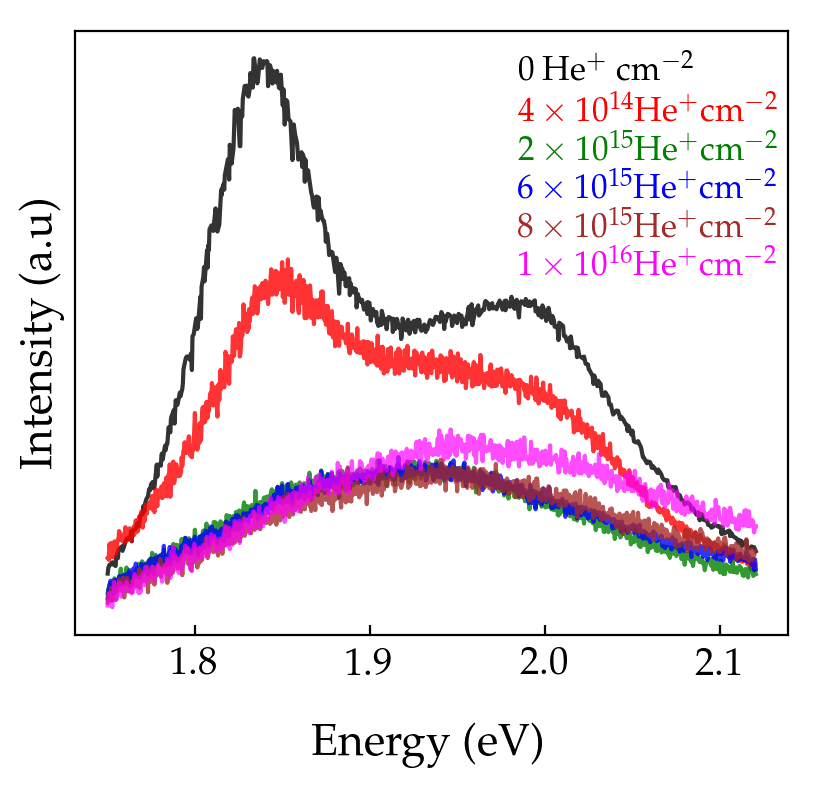}}
\caption{The same data as in Figure 2 but without normalisation, demonstrating the sharp decline in intensity.}
\label{fig:morePL}
\end{figure*}

\FloatBarrier
Figure \ref{fig:maps} (a) is an example of one of the acquired Raman maps of bilayer MoS$_{2}$, presented in this case by summing the intensity of the $A_{1g}$ peak. The bright yellow color represents regions of high intensity. Several darker square regions are visible which were irradiated with different doses of the 30 keV He$^+$ beam. Some other darker regions are monolayer parts of the sample and the darkest regions are the uncovered substrate. Spectra were extracted from sub-regions at least several hundred nanometres from the edges of the flake or the boundaries of the irradiated area and averaged.
Figure \ref{fig:maps} (b) is a PL intensity map of the same region.
Figure \ref{fig:maps} (c) is a white light optical image of the same region where the irradiated regions are only very faintly visible. Figure \ref{fig:morePL} is the same PL data as in Figure 2 but presented without normalisation. 

\clearpage
\section{High Frequency Modes and Dose}
\FloatBarrier

\begin{figure*}
\centering
\subfloat[]{\label{HPosition}\includegraphics[height=2.00in]{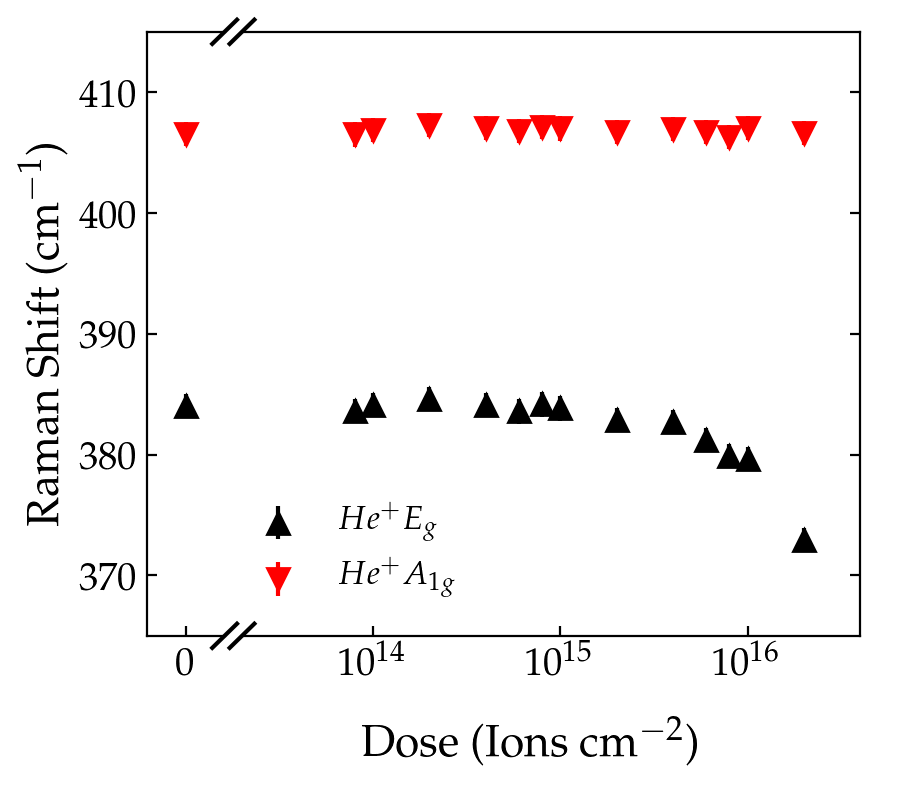}}
\subfloat[]{\label{HWidth}\includegraphics[height=2.00in]{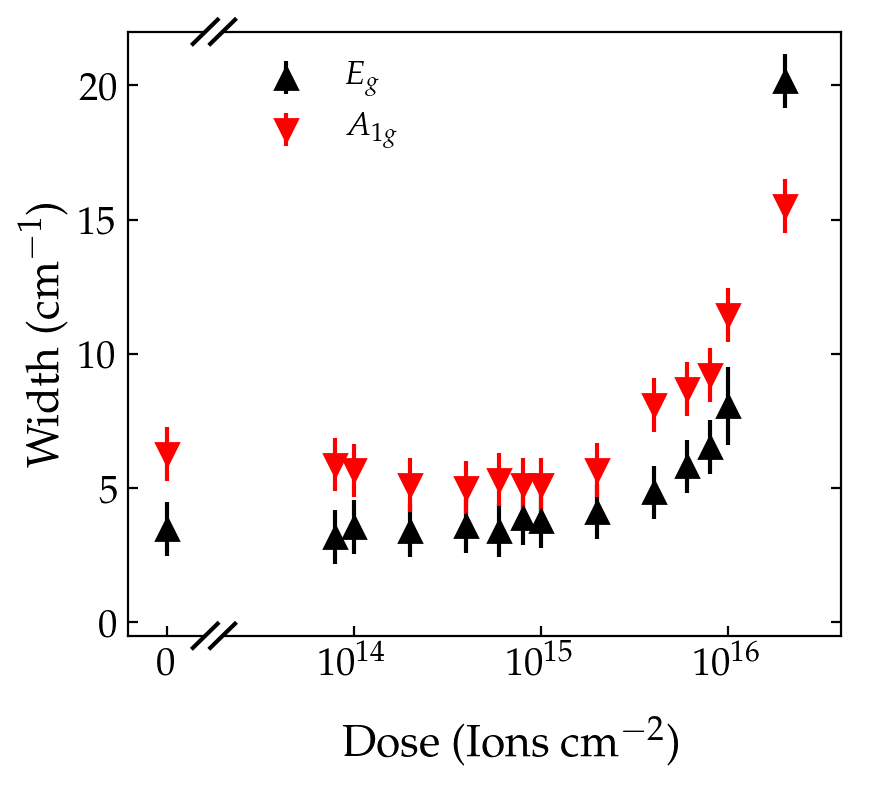}}
\subfloat[]{\label{HIntRatio}\includegraphics[height=2.00in]{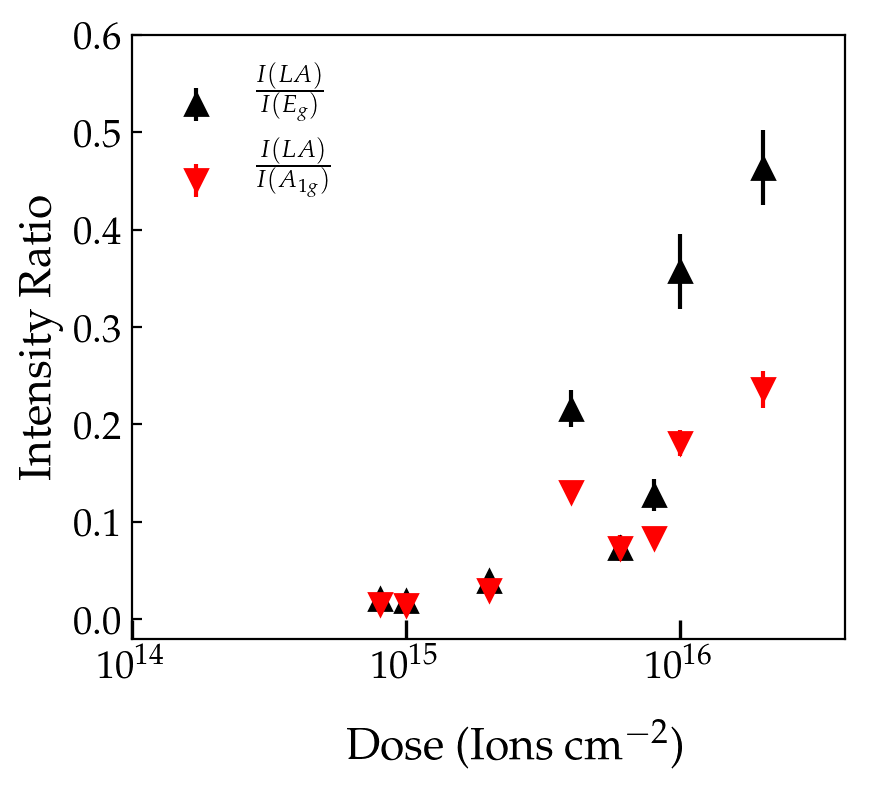}}
\caption{Evolution of the high frequency Raman modes of bilayer MoS$_2$ with increased ion dose.
(a) displays the position of the $A_{1g}$ and $E_{g}$ peaks with increasing dose. With increasing dose, the $E_{g}$ wavenumber decreases significantly. 
(b) displays the increasing width of the $A_{1g}$ and $E_{g}$ peaks with increasing dose.
(c) shows the ratio of intensity of the LA(M) mode to both $A_{1g}$ and $E_{g}$ peaks as a function of dose. The increase in the ratio represents an unambiguous increase in disorder.}
\label{fig:highfrequency}
\end{figure*}

Figure \ref{fig:highfrequency} (a) and (b) show the position and width respectively of the $E_{g}$ and $A_{1g}$ peaks as a function of ion dose. There is no clear upshift in the $A_{1g}$ peak compared to that observed for the $A'$ peak in similar experiments in monolayer MoS$_2$ \cite{Maguire2018}. This may be caused by the bottom layer being protected from exposure to atmosphere (adsorption at defect sites can cause a blue shift). By contrast, the $E_{g}$ red shifts quite considerably for the higher doses, exhibiting very similar results to the $E'$ peak in monolayer MoS$_2$ \cite{Maguire2018}. Both the $E_{g}$ and $A_{1g}$ peaks increase in width as the ion dose increases as expected. 

Figure \ref{fig:highfrequency} (c) shows the increasing intensity of the $LA(M)$ peak normalized to both the $E_{g}$ and $A_{1g}$ peaks as a function of dose. As established by \citeauthor{Mignuzzi2015}, this represents an unambivalent decrease in the average distance between defects \cite{Mignuzzi2015}.

\clearpage
\section{Raman Spectroscopy and Fitting}
\FloatBarrier
\begin{figure*}
\centering
\subfloat[]{\label{HeSpectraFitOne}\includegraphics[height=3.00in]{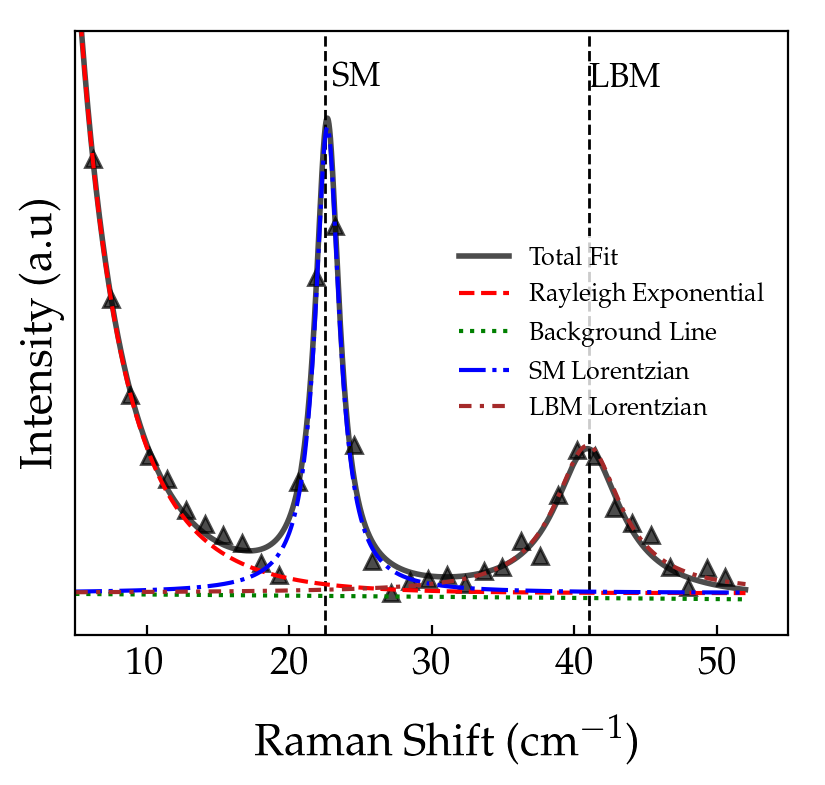}}
\subfloat[]{\label{HeSpectraFitTwo}\includegraphics[height=3.00in]{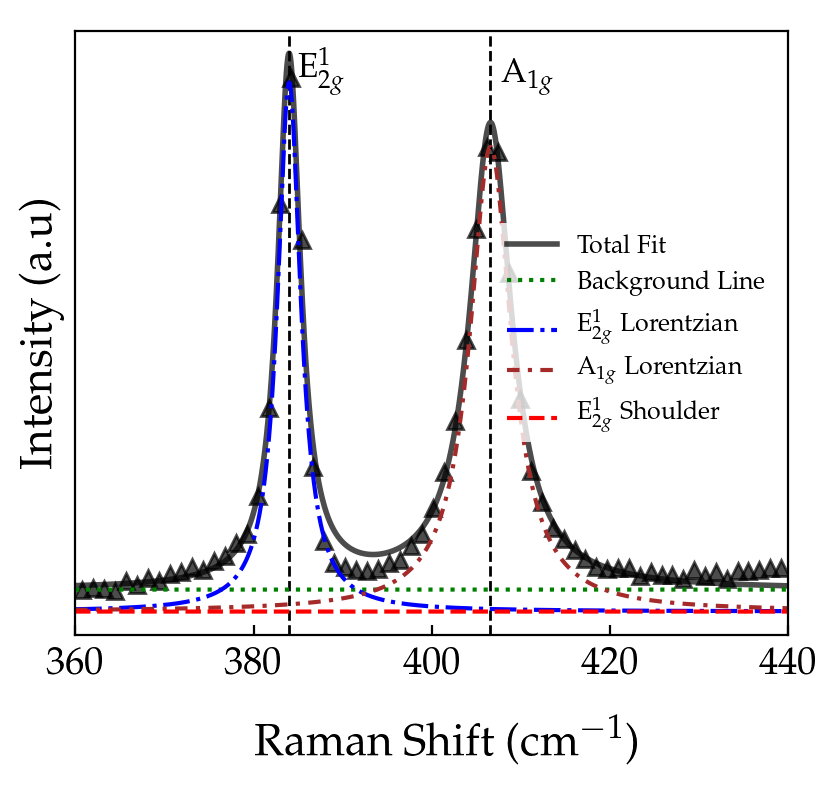}}
\caption{(a) This figure shows the standard fitting process to the SM and LBM. An exponential was fit to the Rayleigh peak, a line was fit to the background and two Lorentzians were fit to the peaks. (b) This figure shows the standard fitting process to the $E_{g}$ and $A_{1g}$ peaks, a line was fit to the background and two Lorentzians were fit to the peaks. An additional Lorentzian was fit to shoulder below the $E_{g}$ peak.}
\label{fig:HeSpectraFitOne}
\end{figure*}

The fitting was performed using the \textit{curve\_{}fit} function in the \textit{scipy} package in python. The bounds and seed values are provided in tables \ref{table:fittingl} and \ref{table:fitting2} for the low and high ranges respectively. Illustrations are provided in Figure \ref{fig:HeSpectraFitOne}. Error bars are from the fitting error. For the low frequency part, points below 5 cm$^{-1}$ and above 51 cm$^{-1}$ were excluded.  The total equation used to fit to the low frequency part of the spectra is given by
\begin{equation}
\begin{aligned}
\underbrace{\color{red}{I_{0}e^{(-\omega^d)}}}_{\text{Rayleigh}} \color{black}{-} \underbrace{\color{green}{m \omega +} \color{green}{c}}_{\text{Baseline}} \color{black}{+} \underbrace{\color{blue}{\frac{I_{SM}}{1+((\omega - \omega _{SM})/\Gamma_{SM})^2}}}_{\text{Shear Mode}}\color{black}{+}\underbrace{\color{Brown}{\frac{I_{LBM}}{1+((\omega - \omega _{LBM})/\Gamma_{LBM})^2}}}_{\text{Layer Breathing Mode}}
\end{aligned}
\end{equation}
Where
$\omega$ is the wavenumber (on the x-axis), 
$I_{0}$ is the intensity scale of the Rayleigh peak,
$d$ is related to the decay of the Rayleigh peak,
$m$ is the slope of the baseline,
$c$ is the is the constant of the baseline,
$I_{SM}$ is the intensity of the SM,
$\omega _{SM}$ is the frequency about which the SM is centred,
$\Gamma_{SM}$ is the HWHM of the SM,
$I_{LBM}$ is the intensity of the LBM,
$\omega_{LBM}$ is the frequency about which the LBM is centred
and $\Gamma_{LBM}$ is the HWHM of the LBM.

For the high frequency part, points below 360 cm$^{-1}$ and above 440 cm$^{-1}$ were excluded. The total equation used to fit to the high frequency part of the spectra is given by

\begin{equation}
\begin{aligned}
\color{black}{-} \underbrace{\color{green}{m \omega +}\color{green}{c}}_{\text{Baseline}} \color{black}{+}  \underbrace{\color{red}{\frac{I_{S}}{1+((\omega - \omega _{S})/\Gamma_{S})^2}}}_{\text{Shoulder}} \color{black}{+} \underbrace{\color{blue}{\frac{I_{E_{g}}}{1+((\omega - \omega _{E_{g}})/\Gamma_{E_{g}})^2}}}_{\text{$E_{g}$}} \color{black}{+} \underbrace{\color{Brown}{\frac{I_{A_{1g}}}{1+((\omega - \omega _{A_{1g}})/\Gamma_{A_{1g}})^2}}}_{\text{$A_{1g}$}}
\end{aligned}
\end{equation}

Where
$m$ is the slope of the baseline, 
$\omega$ is the wavenumber (on the x-axis), 
$c$ is the is the constant of the baseline,
$I_{S}$ is the intensity scale of the shoulder peak,
$\omega _{S}$ is the frequency about which the shoulder peak is centred,
$\Gamma_{S}$ is the HWHM of the shoulder peak,
$I_{E_{g}}$ is the intensity of the $E_{g}$ peak,
$\omega _{E_{g}}$ is the frequency about which the $E_{g}$ peak is centred,
$\Gamma_{E_{g}}$ is the HWHM of the $E_{g}$ peak,
$I_{A_{1g}}$ is the intensity of the $A_{1g}$ peak,
$\omega_{A_{1g}}$ is the frequency about which the $A_{1g}$ peak is centred
and $\Gamma_{A_{1g}}$ is the HWHM of the $A_{1g}$ peak.

\begin{table}[ht]
    \centering
    \begin{tabular}{ c c c c c c c c}
     Labels    & Limit&      &Initial &      & Limit & Unit\\ \hline
     $I_{0}$        & 0   &$\leq$ &$4$   &$\leq$&  10  & a.u. \\
     $d$            & 0   &$\leq$ &$0.59$&$\leq$&  0.7 & \\
     $m$            &-1E-4&$\leq$ &$0$   &$\leq$&  100 &\\
     $c$            &-1E-4&$\leq$ &$0$   &$\leq$&  100 & a.u.\\
     $I_{SM}$       & 0   &$\leq$ &$1$   &$\leq$&  1   & a.u.\\
     $\Gamma_{SM}$  & 0   &$\leq$ &$2.0$ &$\leq$&  100 & cm$^{-1}$\\
     $\omega _{SM}$ & 0   &$\leq$ &$22.8$&$\leq$&  100 & cm$^{-1}$\\
	 $I_{LBM}$      & 0   &$\leq$ &$1$   &$\leq$&  1   & a.u.\\
     $\Gamma_{LBM}$ & 1  &$\leq$ &$3.5 $&$\leq$ &  10  & cm$^{-1}$\\
     $\omega _{LBM}$& 15   &$\leq$ &$41.0$&$\leq$&  45 & cm$^{-1}$\\
     &
    \end{tabular}
    \caption{Initial fitting parameters and their limits for fitting the low frequency range.}
    \vspace{0ex}
    \label{table:fittingl}
\end{table}

\begin{table}[ht]
    \centering
    \begin{tabular}{ c c c c c c c c}
     Labels    & Limit&      &Initial &      & Limit & Unit\\ \hline
     $m$                 & 0   &$\leq$ &$4$   &$\leq$&  0.01 & \\
     $c$                 & 0   &$\leq$ &$0.59$&$\leq$&  100  & a.u.\\
     $I_{S}$             & 0   &$\leq$ &$0$   &$\leq$&  200  & a.u.\\
     $\Gamma_{S}$        & 1   &$\leq$ &$0$   &$\leq$&  100  & cm$^{-1}$\\
     $\omega _{S}$       & 200 &$\leq$ &$1$   &$\leq$&  400  & cm$^{-1}$\\
     $I_{E_{g}}$      & 0   &$\leq$ &$2.0$ &$\leq$&  300  & a.u.\\
     $\Gamma _{E_{g}}$& 1   &$\leq$ &$22.8$&$\leq$&  100  & cm$^{-1}$\\
	 $\omega_{E_{g}}$ & 200 &$\leq$ &$1$   &$\leq$&  460  & cm$^{-1}$\\
     $I_{A_{1g}}$        & 0   &$\leq$ &$3.5 $&$\leq$ & 300  & a.u.\\
     $\Gamma_{A_{1g}}$   & 1   &$\leq$ &$41.0$&$\leq$&  100  & cm$^{-1}$\\
     $\omega_{A_{1g}}$   & 200 &$\leq$ &$41.0$&$\leq$&  380  & cm$^{-1}$\\
     &
    \end{tabular}
    \caption{Initial fitting parameters and their limits for fitting the high frequency range.}
    \vspace{0ex}
    \label{table:fitting2}
\end{table}
\clearpage

\section{Lamella Preparation}
\FloatBarrier

\begin{figure*}
\centering
\subfloat[]{\label{lamellaoptical}\includegraphics[height=2.2in]{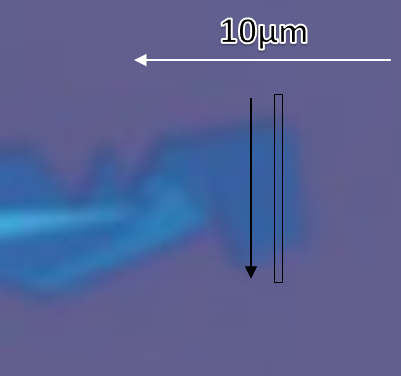}}\quad
\subfloat[]{\label{lamellaschematic}\includegraphics[height=2.2in]{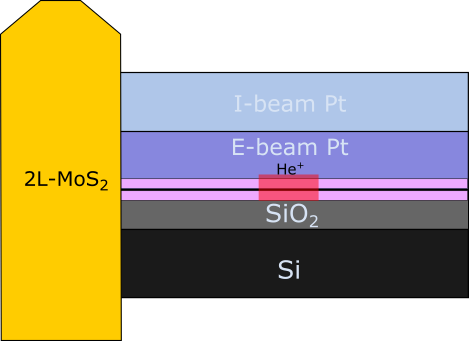}}\quad
\caption{(a) is an optical image showing the flake of mechanically exfoliated MoS$_2$ used in TEM experiments. The black box approximately indicates the region irradiated with 6$\times$ 10$^{15}$ He$^+$ cm$^{-2}$. (b) is a schematic of the lamella which is not drawn to scale.} 
\label{fig:lamellaprep}
\end{figure*}

MoS$_2$ for electron microscopy was mechanically exfoliated using the adhesive tape method from commercially available bulk molybdenite crystals (\textit{SPI Supplies}). This method was chosen for the low level of pre-existing defects. Suitable flakes were identified and irradiated in similar fashion to the Raman experiments as shown in the optical image of Figure \ref{fig:lamellaprep} (a). 
The flakes were deposited on a 285 nm layer of SiO$_2$ on Si and bilayer regions with straight edges were identified using optical microscopy. A suitable flake was then irradiated in the \textit{Zeiss ORION Nanofab} with a dose of 6 $\times$ $10^{15}$ He cm$^{-2}$. The pixel spacing was 5 nm and the beam current was $\sim$1 pA.

To prepare the cross-sectional lamella, the sample was loaded into a \textit{Zeiss Auriga} focused ion beam system. The surface was covered locally with several hundred nanometres of a protective platinum-based coating, first using an electron beam at 5 keV and then using the 30 keV Ga$^+$ beam. The lamella was cut using the Ga$^+$ ion beam at 30 keV and lifted out using a nanomanipulator needle and welded to a copper TEM grid. Initial thinning was performed with a 15 keV Ga$^+$ beam. 

The sample was then cleaned for 90s in a \textit{Fischione Instruments 1020} plasma cleaner which uses a 1:3 mixture of O$_2$/Ar gas.
The sample was further thinned using a \textit{Fischione 1040 Nanomill}. Argon ions at 900 eV and a beam current of 100 pA were incident at an angle of $\sim$70 degrees from the bottom (Si) of the lamella. The thickness was checked between each irradiation in the TEM. One side was irradiated for 13 mins total and the other for 4 mins total. The finished lamella is illustrated in \ref{fig:lamellaprep} (b). 
 
Electron microscopy was then carried out on the mechanically exfoliated bilayer MoS$_2$ in a \textit{FEI Titan 80-300} operated at 300 keV with a chamber pressure of $\sim$4 $\times$ 10$^{-8}$ mbar.


\clearpage

\section{Layer Separation Measurement}
\begin{figure*}
\centering
\subfloat{\label{sub:individuallineprofile}\includegraphics[height=2.2in]{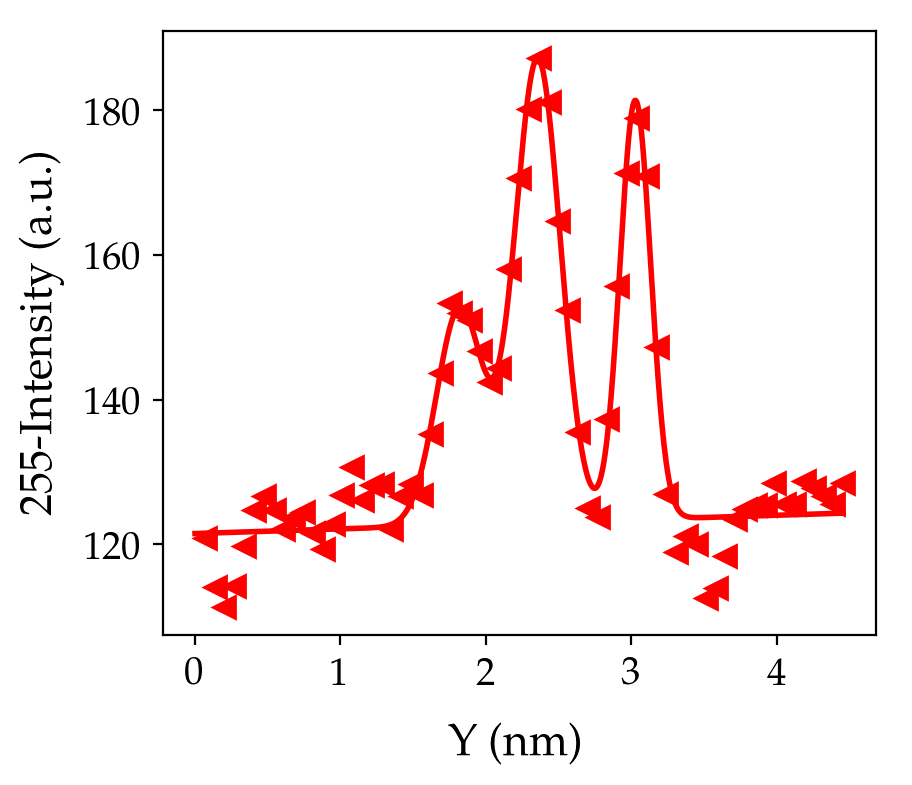}}\quad
\caption{Example of a line profile across the MoS$_2$ layers in the TEM images in the main test. Fitted peaks allowed the extraction of the histograms in the main text. } 
\label{fig:lineprofiles}
\end{figure*}

\FloatBarrier
A \textit{python} script was written which does the following:
\begin{itemize}
\item reads the columns of a TEM image (such as those in the main text) as line profiles and averages them over 2 pixels to each side to reduce noise.
\item fits three Gaussian distributions to each line profile (as demonstrated in Figure \ref{fig:lineprofiles}) corresponding to the bottom layer, top layer and shoulder to the top layer.
\item calculates the separation of the two layers and plots them in a histogram as in the main text.
\item where the fitting failed or was poor (R$^2<0.75$) then that point was ignored. 
\end{itemize}

\FloatBarrier
\clearpage

\section{Sample Preparation and Irradiation}
\FloatBarrier

MoS$_2$ was prepared as follows: MoO$_3$ substrates were placed face-up in a ceramic boat with a blank SiO$_2$ substrate face-down on top. This was situated in the centre of the heating zone of a quartz tube furnace, and ramped to 750 $^{\circ}$C under 150 SCCM of Ar flow. Sulfur (S) vapour was then produced by heating S powder to $\sim$120 $^{\circ}$C in an independently controlled upstream heating zone of the furnace, and carried downstream to the MoO$_3$ for a duration of 20 min. After this, the furnace was held at 750~$^{\circ}$C for 20 min, then cooled down to room temperature. 
Flakes of MoS$_2$ with desired thickness were identified on the SiO$_2$ surface by optical contrast and Raman spectroscopy \cite{Lee2010, Gomez2012}.

The flakes were then irradiated as described in the main text and Raman spectra of irradiated regions collected. Irradiated patterns in the initial Raman experiments (Figures 1,2 \& 3) have a pixel spacing of 10 nm and a size of $4 \times 4$ $\mu m^2$. Patterns prepared for HRTEM have a pixel spacing of 5 nm (Figure 4). The ion beam was defocussed ($\sim$10s of nm) to ensure a uniform distribution of ions. The pixel spacing was then varied for the final Raman experiment with a focused probe (Figure 6) and a size of $3 \times 3$ $\mu m^2$. A beam current of $\sim$ 1 pA was used throughout. The beam dwell time at each pixel and/or the number of repeats at each position were varied to achieve the desired dose. The chamber pressure was of the order 3 $\times$ 10$^{-7}$ Torr. 

To measure the probe size, a helium ion image was acquired of a nearby high contrast edge on the sample surface shown in Figure S6. The probe size was measured to be 4.0 $\pm$ 0.9 nm using the \textit{Gauss Fit} plug-in in \textit{Imagej} \cite{Rueden2017}.
It was assumed that the feature being imaged had a perfectly sharp edge. The decline in secondary electron intensity over the edge in the image is then
attributed to the distribution of the ions in the probe \cite{Fox2015}. 
This intensity profile was plotted and the distance between 25\% and 75\% of the average intensity of the flake was taken as the probe size. This is equivalent to the full width at half maximum of the probe.
\begin{figure*}
\centering
\subfloat[]{\label{sub:HeImage}\includegraphics[height=2.00in]{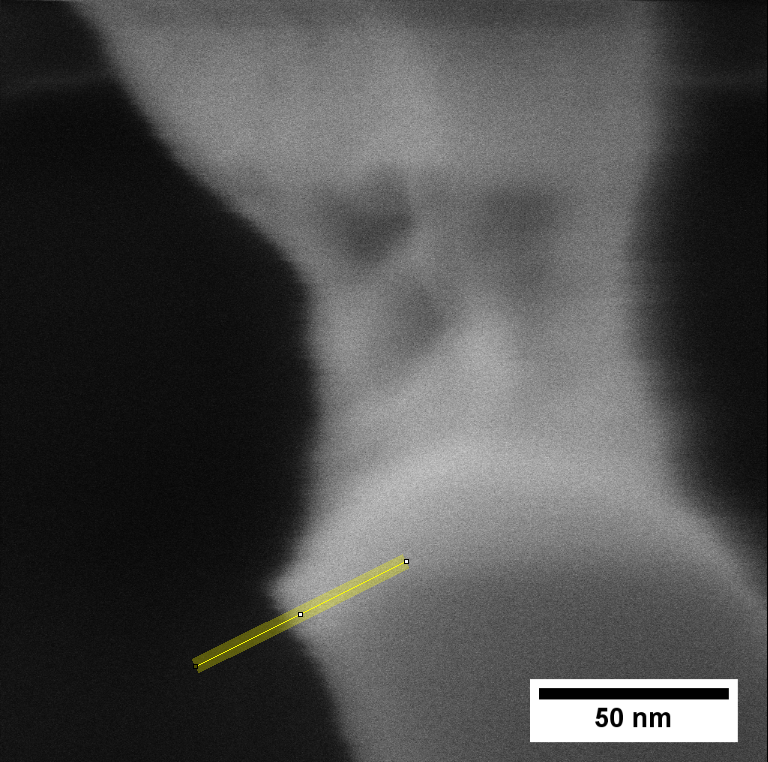}}\quad
\subfloat[]{\label{sub:HeProbe}\includegraphics[height=1.50in]{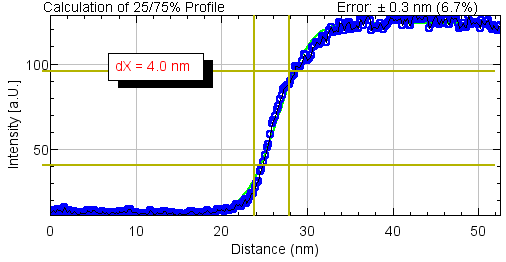}}
\caption{Measuring the probe size of the helium ion beam on a high contrast feature near the irradiation zone. (a) is a secondary electron image from the helium ion microscope with a yellow line showing where the profile was extracted from. (b) shows the extracted line profile and the result of the probe size calculation performed in \textit{Imagej} using the \textit{Gauss Fit} plug-in}
\label{fig:HeProbe}
\end{figure*}
\clearpage

\section{Mass Loss}
\FloatBarrier
It has been previously reported that electron and ion irradiation of MoS$_2$ can lead to the preferential sputtering of sulfur atoms \cite{Fox2015, Parkin2016}. The resulting changes to the Raman spectrum were studied in detail by \citeauthor{Parkin2016} in monolayer MoS$_2$. We applied the assumption that the difference in changes to the peaks of 1L and 2L MoS$_2$ would be negligible. We used their results to estimate the percentage of S atoms lost due to He$^+$ irradiation in this work based on the shift of the $E_{g}$ peak and the increase in peak separation. 

\begin{figure*}
\centering
\subfloat[]{\label{sub:sulfurvacancies}\includegraphics[height=2.4in]{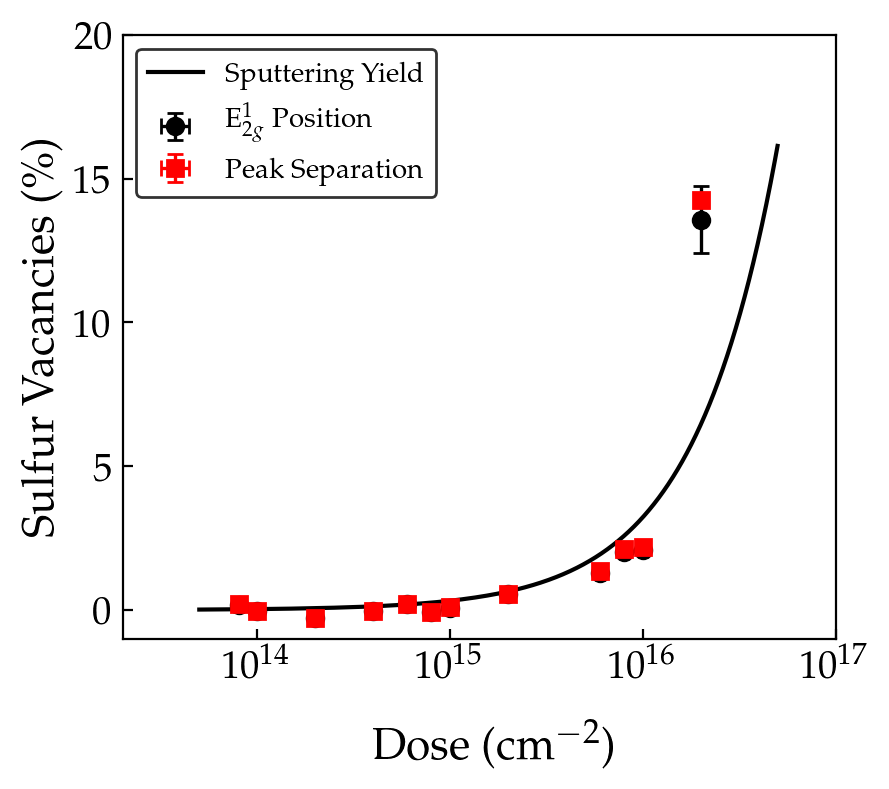}}\quad
\caption{Calculation of mass lost based on comparison of Raman spectra to literature (points) \protect \cite{Parkin2016} and knowledge of sputtering yield (line)\protect \cite{Maguire2018,Kretschmer2018}} 
\label{fig:sulfurvacancies}
\end{figure*}

Figure \ref{fig:sulfurvacancies} shows the missing sulfur percentage as a function of dose. The black markers are estimates from $E_{g}$ peak shift, the red are from peak separation changes and the line is from the sputtering yield in the two cited papers \cite{Maguire2018,Kretschmer2018}.

The ion beam is expected to remove material with a preference for S atoms and S is heavier than any of the atmospheric species likely to fill its vacancies, decreasing $\mu$ with increasing dose. By comparison to literature \cite{Parkin2016} we see that changes to $\mu$ are small except at the highest doses and from the mass-adjusted LCM equation in the main paper those changes to $\mu$ are expected to cause a blue-shift, not red. Therefore we dismiss changes in $\mu$ as a cause for changing $\omega$. 

\FloatBarrier
\clearpage

\section{Calculated Layer Separation}
\FloatBarrier
From the mass-adjusted LCM equation in the main paper we can calculate the increase in interlayer separation which would be required to explain the red shift if we exclude twisting and shear modulus effects. If the red shift of the shear mode was to be explained within the LCM by changes only to layer separation, then it would be calculated as follows:

\begin{equation}
t = \frac{1}{\sqrt[]{2}\pi ^2 c^2} \frac{C}{\mu_0 (1 - \frac{\gamma S}{n})\omega ^2}(1 + \cos(\frac{\pi}{N}))
\end{equation}
where $t$ is the interlayer separation, 
$c$ is the speed of light in cm s$^{-1}$, 
$C$ is the shear modulus, 
$\mu_0$ is the single layer mass per unit area of pristine MoS$_2$,
$\gamma$ is the sputter yield,
$S$, is the dose,
$n$ is the atomic density, 
$\omega$ is the experimentally obtained change in the frequency of the shear mode,
$N$ is the layer number (fixed at 2),

For direct comparison to our TEM experiment we calculate that for a dose of 6$\times$10$^{15}$ He$^+$ cm$^{-2}$, an increase in interlayer separation of $\sim$0.11 nm would be required. Since the results of this calculation do not agree with the TEM results presented in the main paper, we conclude that the change in layer separation is negligible and instead that twisting and shear modulus effects are responsible for the shift.

\begin{figure*}
\centering
\subfloat[]{\label{sub:Llayerseparation}\includegraphics[height=2.4in]{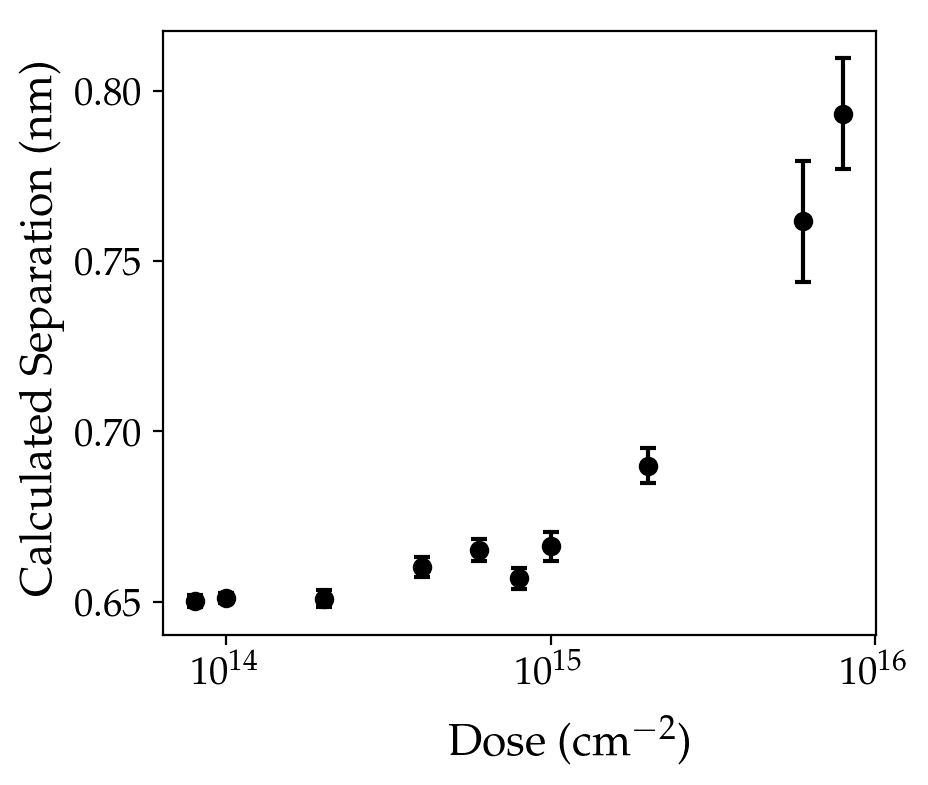}}\quad
\caption{(a) shows the calculated layer separation based on changes to the shear mode (the results of which do not agree with TEM results in the main paper). 
} 
\label{fig:Llayerseparation}
\end{figure*}
\FloatBarrier
\clearpage

\section{Twisting: Intensity and Frequency}
\begin{figure*}
\centering
\subfloat[]{\label{sub:LseparationIntRWL}\includegraphics[height=2.4in]{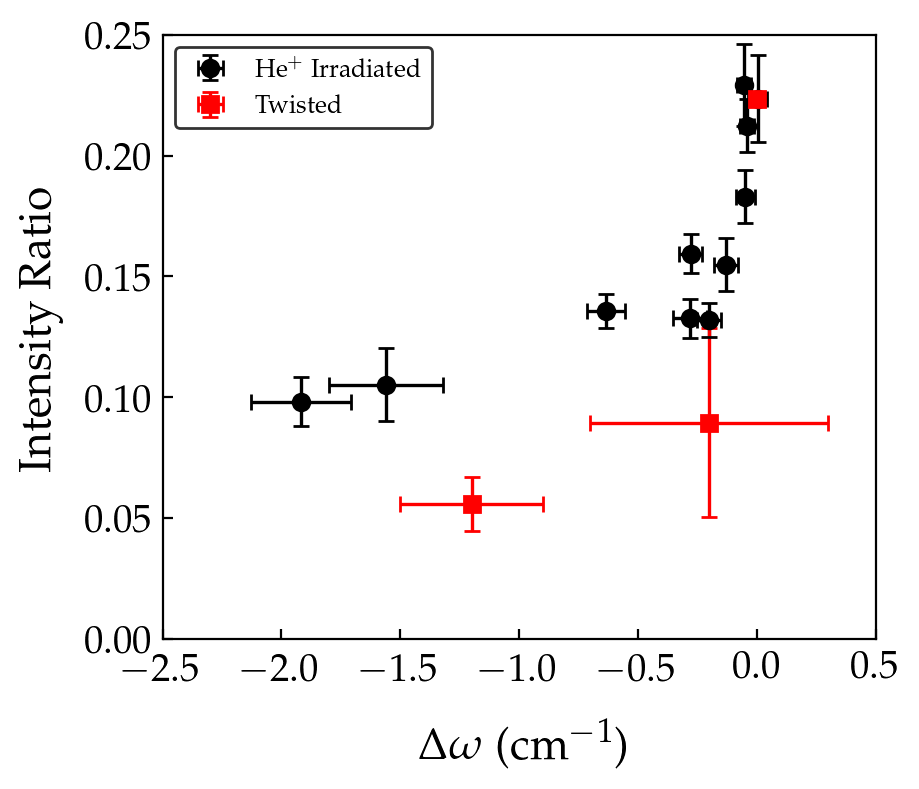}}\quad
\caption{A comparison of our irradiated 2L MoS$_2$ data (black) to the twisted bilayers of \citeauthor{Huang2016} (red) \protect \cite{Huang2016}. } 
\label{fig:LseparationIntRWL}
\end{figure*}

\FloatBarrier
Figure \ref{fig:LseparationIntRWL} is a comparison of our data to that of \citeauthor{Huang2016} \cite{Huang2016}. Given the similarity, this figure shows that twisting is likely playing a large role in our irradiated material (particularly in explaining the loss of SM intensity) but orientation/twisting effects do not fully explain the red shift observed at a given value of the intensity ratio. Therefore, we suggest that reduced crystallinity causes a change in the shear modulus. 
\FloatBarrier
\clearpage

\section{Bibliography}
\FloatBarrier
\twocolumngrid

\bibliography{library}

\begin{thebibliography}{42}%
\makeatletter
\providecommand \@ifxundefined [1]{%
 \@ifx{#1\undefined}
}%
\providecommand \@ifnum [1]{%
 \ifnum #1\expandafter \@firstoftwo
 \else \expandafter \@secondoftwo
 \fi
}%
\providecommand \@ifx [1]{%
 \ifx #1\expandafter \@firstoftwo
 \else \expandafter \@secondoftwo
 \fi
}%
\providecommand \natexlab [1]{#1}%
\providecommand \enquote  [1]{``#1''}%
\providecommand \bibnamefont  [1]{#1}%
\providecommand \bibfnamefont [1]{#1}%
\providecommand \citenamefont [1]{#1}%
\providecommand \href@noop [0]{\@secondoftwo}%
\providecommand \href [0]{\begingroup \@sanitize@url \@href}%
\providecommand \@href[1]{\@@startlink{#1}\@@href}%
\providecommand \@@href[1]{\endgroup#1\@@endlink}%
\providecommand \@sanitize@url [0]{\catcode `\\12\catcode `\$12\catcode
  `\&12\catcode `\#12\catcode `\^12\catcode `\_12\catcode `\%12\relax}%
\providecommand \@@startlink[1]{}%
\providecommand \@@endlink[0]{}%
\providecommand \url  [0]{\begingroup\@sanitize@url \@url }%
\providecommand \@url [1]{\endgroup\@href {#1}{\urlprefix }}%
\providecommand \urlprefix  [0]{URL }%
\providecommand \Eprint [0]{\href }%
\providecommand \doibase [0]{http://dx.doi.org/}%
\providecommand \selectlanguage [0]{\@gobble}%
\providecommand \bibinfo  [0]{\@secondoftwo}%
\providecommand \bibfield  [0]{\@secondoftwo}%
\providecommand \translation [1]{[#1]}%
\providecommand \BibitemOpen [0]{}%
\providecommand \bibitemStop [0]{}%
\providecommand \bibitemNoStop [0]{.\EOS\space}%
\providecommand \EOS [0]{\spacefactor3000\relax}%
\providecommand \BibitemShut  [1]{\csname bibitem#1\endcsname}%
\let\auto@bib@innerbib\@empty
\bibitem [{\citenamefont {Novoselov}\ \emph {et~al.}(2016)\citenamefont
  {Novoselov}, \citenamefont {Mishchenko}, \citenamefont {Carvalho},\ and\
  \citenamefont {{Castro Neto}}}]{Novoselov2016}%
  \BibitemOpen
  \bibfield  {author} {\bibinfo {author} {\bibfnamefont {K.~S.}\ \bibnamefont
  {Novoselov}}, \bibinfo {author} {\bibfnamefont {A.}~\bibnamefont
  {Mishchenko}}, \bibinfo {author} {\bibfnamefont {A.}~\bibnamefont
  {Carvalho}}, \ and\ \bibinfo {author} {\bibfnamefont {A.~H.}\ \bibnamefont
  {{Castro Neto}}},\ }\href {\doibase 10.1126/science.aac9439} {\bibfield
  {journal} {\bibinfo  {journal} {Science}\ }\textbf {\bibinfo {volume}
  {353}},\ \bibinfo {pages} {aac9439} (\bibinfo {year} {2016})}\BibitemShut
  {NoStop}%
\bibitem [{\citenamefont {Jariwala}\ \emph {et~al.}(2016)\citenamefont
  {Jariwala}, \citenamefont {Marks},\ and\ \citenamefont
  {Hersam}}]{Jariwala2016}%
  \BibitemOpen
  \bibfield  {author} {\bibinfo {author} {\bibfnamefont {D.}~\bibnamefont
  {Jariwala}}, \bibinfo {author} {\bibfnamefont {T.~J.}\ \bibnamefont {Marks}},
  \ and\ \bibinfo {author} {\bibfnamefont {M.~C.}\ \bibnamefont {Hersam}},\
  }\href {\doibase 10.1038/nmat4703} {\bibfield  {journal} {\bibinfo  {journal}
  {Nature Materials}\ }\textbf {\bibinfo {volume} {16}},\ \bibinfo {pages}
  {170} (\bibinfo {year} {2016})}\BibitemShut {NoStop}%
\bibitem [{\citenamefont {Kang}\ \emph {et~al.}(2017)\citenamefont {Kang},
  \citenamefont {Lee}, \citenamefont {Han}, \citenamefont {Gao}, \citenamefont
  {Xie}, \citenamefont {Muller},\ and\ \citenamefont {Park}}]{Kang2017}%
  \BibitemOpen
  \bibfield  {author} {\bibinfo {author} {\bibfnamefont {K.}~\bibnamefont
  {Kang}}, \bibinfo {author} {\bibfnamefont {K.~H.}\ \bibnamefont {Lee}},
  \bibinfo {author} {\bibfnamefont {Y.}~\bibnamefont {Han}}, \bibinfo {author}
  {\bibfnamefont {H.}~\bibnamefont {Gao}}, \bibinfo {author} {\bibfnamefont
  {S.}~\bibnamefont {Xie}}, \bibinfo {author} {\bibfnamefont {D.~A.}\
  \bibnamefont {Muller}}, \ and\ \bibinfo {author} {\bibfnamefont
  {J.}~\bibnamefont {Park}},\ }\href {\doibase 10.1038/nature23905} {\bibfield
  {journal} {\bibinfo  {journal} {Nature}\ }\textbf {\bibinfo {volume} {550}},\
  \bibinfo {pages} {229} (\bibinfo {year} {2017})}\BibitemShut {NoStop}%
\bibitem [{\citenamefont {Sung}\ \emph {et~al.}(2017)\citenamefont {Sung},
  \citenamefont {Heo}, \citenamefont {Si}, \citenamefont {Kim}, \citenamefont
  {Noh}, \citenamefont {Song}, \citenamefont {Kim}, \citenamefont {Lee},
  \citenamefont {Seo}, \citenamefont {Kim}, \citenamefont {Kim}, \citenamefont
  {Yeom}, \citenamefont {Kim}, \citenamefont {Choi}, \citenamefont {Kim},\ and\
  \citenamefont {Jo}}]{Sung2017}%
  \BibitemOpen
  \bibfield  {author} {\bibinfo {author} {\bibfnamefont {J.~H.}\ \bibnamefont
  {Sung}}, \bibinfo {author} {\bibfnamefont {H.}~\bibnamefont {Heo}}, \bibinfo
  {author} {\bibfnamefont {S.}~\bibnamefont {Si}}, \bibinfo {author}
  {\bibfnamefont {Y.~H.}\ \bibnamefont {Kim}}, \bibinfo {author} {\bibfnamefont
  {H.~R.}\ \bibnamefont {Noh}}, \bibinfo {author} {\bibfnamefont
  {K.}~\bibnamefont {Song}}, \bibinfo {author} {\bibfnamefont {J.}~\bibnamefont
  {Kim}}, \bibinfo {author} {\bibfnamefont {C.~S.}\ \bibnamefont {Lee}},
  \bibinfo {author} {\bibfnamefont {S.~Y.}\ \bibnamefont {Seo}}, \bibinfo
  {author} {\bibfnamefont {D.~H.}\ \bibnamefont {Kim}}, \bibinfo {author}
  {\bibfnamefont {H.~K.}\ \bibnamefont {Kim}}, \bibinfo {author} {\bibfnamefont
  {H.~W.}\ \bibnamefont {Yeom}}, \bibinfo {author} {\bibfnamefont {T.~H.}\
  \bibnamefont {Kim}}, \bibinfo {author} {\bibfnamefont {S.}~\bibnamefont
  {Choi}}, \bibinfo {author} {\bibfnamefont {J.~S.}\ \bibnamefont {Kim}}, \
  and\ \bibinfo {author} {\bibfnamefont {M.~H.}\ \bibnamefont {Jo}},\ }\href
  {\doibase 10.1038/nnano.2017.161} {\bibfield  {journal} {\bibinfo  {journal}
  {Nature Nanotechnology}\ }\textbf {\bibinfo {volume} {12}},\ \bibinfo {pages}
  {1064} (\bibinfo {year} {2017})}\BibitemShut {NoStop}%
\bibitem [{\citenamefont {Boukhicha}\ \emph {et~al.}(2013)\citenamefont
  {Boukhicha}, \citenamefont {Calandra}, \citenamefont {Measson}, \citenamefont
  {Lancry},\ and\ \citenamefont {Shukla}}]{Boukhicha2013}%
  \BibitemOpen
  \bibfield  {author} {\bibinfo {author} {\bibfnamefont {M.}~\bibnamefont
  {Boukhicha}}, \bibinfo {author} {\bibfnamefont {M.}~\bibnamefont {Calandra}},
  \bibinfo {author} {\bibfnamefont {M.~A.}\ \bibnamefont {Measson}}, \bibinfo
  {author} {\bibfnamefont {O.}~\bibnamefont {Lancry}}, \ and\ \bibinfo {author}
  {\bibfnamefont {A.}~\bibnamefont {Shukla}},\ }\href {\doibase
  10.1103/PhysRevB.87.195316} {\bibfield  {journal} {\bibinfo  {journal}
  {Physical Review B}\ }\textbf {\bibinfo {volume} {87}},\ \bibinfo {pages} {1}
  (\bibinfo {year} {2013})}\BibitemShut {NoStop}%
\bibitem [{\citenamefont {Lui}\ \emph {et~al.}(2015)\citenamefont {Lui},
  \citenamefont {Ye}, \citenamefont {Ji}, \citenamefont {Chiu}, \citenamefont
  {Chou}, \citenamefont {Andersen}, \citenamefont {Means-Shively},
  \citenamefont {Anderson}, \citenamefont {Wu}, \citenamefont {Kidd},
  \citenamefont {Lee},\ and\ \citenamefont {He}}]{Lui2015}%
  \BibitemOpen
  \bibfield  {author} {\bibinfo {author} {\bibfnamefont {C.~H.}\ \bibnamefont
  {Lui}}, \bibinfo {author} {\bibfnamefont {Z.}~\bibnamefont {Ye}}, \bibinfo
  {author} {\bibfnamefont {C.}~\bibnamefont {Ji}}, \bibinfo {author}
  {\bibfnamefont {K.~C.}\ \bibnamefont {Chiu}}, \bibinfo {author}
  {\bibfnamefont {C.~T.}\ \bibnamefont {Chou}}, \bibinfo {author}
  {\bibfnamefont {T.~I.}\ \bibnamefont {Andersen}}, \bibinfo {author}
  {\bibfnamefont {C.}~\bibnamefont {Means-Shively}}, \bibinfo {author}
  {\bibfnamefont {H.}~\bibnamefont {Anderson}}, \bibinfo {author}
  {\bibfnamefont {J.~M.}\ \bibnamefont {Wu}}, \bibinfo {author} {\bibfnamefont
  {T.}~\bibnamefont {Kidd}}, \bibinfo {author} {\bibfnamefont {Y.~H.}\
  \bibnamefont {Lee}}, \ and\ \bibinfo {author} {\bibfnamefont
  {R.}~\bibnamefont {He}},\ }\href {\doibase 10.1103/PhysRevB.91.165403}
  {\bibfield  {journal} {\bibinfo  {journal} {Physical Review B}\ }\textbf
  {\bibinfo {volume} {91}},\ \bibinfo {pages} {1} (\bibinfo {year}
  {2015})}\BibitemShut {NoStop}%
\bibitem [{\citenamefont {Zhang}\ \emph {et~al.}(2016)\citenamefont {Zhang},
  \citenamefont {Wang}, \citenamefont {Chen}, \citenamefont {Sun},
  \citenamefont {Wu}, \citenamefont {Jia}, \citenamefont {Lu}, \citenamefont
  {Yu}, \citenamefont {Chen}, \citenamefont {Zhu}, \citenamefont {Xie},
  \citenamefont {Yang}, \citenamefont {Shi}, \citenamefont {Xu}, \citenamefont
  {Xiang}, \citenamefont {Liu},\ and\ \citenamefont {Zhang}}]{Zhang2016a}%
  \BibitemOpen
  \bibfield  {author} {\bibinfo {author} {\bibfnamefont {J.}~\bibnamefont
  {Zhang}}, \bibinfo {author} {\bibfnamefont {J.}~\bibnamefont {Wang}},
  \bibinfo {author} {\bibfnamefont {P.}~\bibnamefont {Chen}}, \bibinfo {author}
  {\bibfnamefont {Y.}~\bibnamefont {Sun}}, \bibinfo {author} {\bibfnamefont
  {S.}~\bibnamefont {Wu}}, \bibinfo {author} {\bibfnamefont {Z.}~\bibnamefont
  {Jia}}, \bibinfo {author} {\bibfnamefont {X.}~\bibnamefont {Lu}}, \bibinfo
  {author} {\bibfnamefont {H.}~\bibnamefont {Yu}}, \bibinfo {author}
  {\bibfnamefont {W.}~\bibnamefont {Chen}}, \bibinfo {author} {\bibfnamefont
  {J.}~\bibnamefont {Zhu}}, \bibinfo {author} {\bibfnamefont {G.}~\bibnamefont
  {Xie}}, \bibinfo {author} {\bibfnamefont {R.}~\bibnamefont {Yang}}, \bibinfo
  {author} {\bibfnamefont {D.}~\bibnamefont {Shi}}, \bibinfo {author}
  {\bibfnamefont {X.}~\bibnamefont {Xu}}, \bibinfo {author} {\bibfnamefont
  {J.}~\bibnamefont {Xiang}}, \bibinfo {author} {\bibfnamefont
  {K.}~\bibnamefont {Liu}}, \ and\ \bibinfo {author} {\bibfnamefont
  {G.}~\bibnamefont {Zhang}},\ }\href {\doibase 10.1002/adma.201504631}
  {\bibfield  {journal} {\bibinfo  {journal} {Advanced Materials}\ }\textbf
  {\bibinfo {volume} {28}},\ \bibinfo {pages} {1950} (\bibinfo {year}
  {2016})}\BibitemShut {NoStop}%
\bibitem [{\citenamefont {Li}\ \emph {et~al.}(2016)\citenamefont {Li},
  \citenamefont {Lin}, \citenamefont {Chui},\ and\ \citenamefont
  {Lau}}]{Li2016}%
  \BibitemOpen
  \bibfield  {author} {\bibinfo {author} {\bibfnamefont {Y.}~\bibnamefont
  {Li}}, \bibinfo {author} {\bibfnamefont {S.}~\bibnamefont {Lin}}, \bibinfo
  {author} {\bibfnamefont {Y.~S.}\ \bibnamefont {Chui}}, \ and\ \bibinfo
  {author} {\bibfnamefont {S.~P.}\ \bibnamefont {Lau}},\ }\href {\doibase
  10.1149/2.0071611jss} {\bibfield  {journal} {\bibinfo  {journal} {ECS Journal
  of Solid State Science and Technology}\ }\textbf {\bibinfo {volume} {5}},\
  \bibinfo {pages} {Q3033} (\bibinfo {year} {2016})}\BibitemShut {NoStop}%
\bibitem [{\citenamefont {Zhao}\ \emph {et~al.}(2013)\citenamefont {Zhao},
  \citenamefont {Luo}, \citenamefont {Li}, \citenamefont {Zhang}, \citenamefont
  {Araujo}, \citenamefont {Gan}, \citenamefont {Wu}, \citenamefont {Zhang},
  \citenamefont {Quek}, \citenamefont {Dresselhaus},\ and\ \citenamefont
  {Xiong}}]{Zhao2013}%
  \BibitemOpen
  \bibfield  {author} {\bibinfo {author} {\bibfnamefont {Y.}~\bibnamefont
  {Zhao}}, \bibinfo {author} {\bibfnamefont {X.}~\bibnamefont {Luo}}, \bibinfo
  {author} {\bibfnamefont {H.}~\bibnamefont {Li}}, \bibinfo {author}
  {\bibfnamefont {J.}~\bibnamefont {Zhang}}, \bibinfo {author} {\bibfnamefont
  {P.~T.}\ \bibnamefont {Araujo}}, \bibinfo {author} {\bibfnamefont {C.~K.}\
  \bibnamefont {Gan}}, \bibinfo {author} {\bibfnamefont {J.}~\bibnamefont
  {Wu}}, \bibinfo {author} {\bibfnamefont {H.}~\bibnamefont {Zhang}}, \bibinfo
  {author} {\bibfnamefont {S.~Y.}\ \bibnamefont {Quek}}, \bibinfo {author}
  {\bibfnamefont {M.~S.}\ \bibnamefont {Dresselhaus}}, \ and\ \bibinfo {author}
  {\bibfnamefont {Q.}~\bibnamefont {Xiong}},\ }\href {\doibase
  10.1021/nl304169w} {\bibfield  {journal} {\bibinfo  {journal} {Nano Letters}\
  }\textbf {\bibinfo {volume} {13}},\ \bibinfo {pages} {1007} (\bibinfo {year}
  {2013})}\BibitemShut {NoStop}%
\bibitem [{\citenamefont {Zhang}\ \emph {et~al.}(2013)\citenamefont {Zhang},
  \citenamefont {Han}, \citenamefont {Wu}, \citenamefont {Milana},
  \citenamefont {Lu}, \citenamefont {Li}, \citenamefont {Ferrari},\ and\
  \citenamefont {Tan}}]{Zhang2013}%
  \BibitemOpen
  \bibfield  {author} {\bibinfo {author} {\bibfnamefont {X.}~\bibnamefont
  {Zhang}}, \bibinfo {author} {\bibfnamefont {W.~P.}\ \bibnamefont {Han}},
  \bibinfo {author} {\bibfnamefont {J.~B.}\ \bibnamefont {Wu}}, \bibinfo
  {author} {\bibfnamefont {S.}~\bibnamefont {Milana}}, \bibinfo {author}
  {\bibfnamefont {Y.}~\bibnamefont {Lu}}, \bibinfo {author} {\bibfnamefont
  {Q.~Q.}\ \bibnamefont {Li}}, \bibinfo {author} {\bibfnamefont {A.~C.}\
  \bibnamefont {Ferrari}}, \ and\ \bibinfo {author} {\bibfnamefont {P.~H.}\
  \bibnamefont {Tan}},\ }\href@noop {} {\bibfield  {journal} {\bibinfo
  {journal} {Physical Review B}\ }\textbf {\bibinfo {volume} {87}},\ \bibinfo
  {pages} {115413} (\bibinfo {year} {2013})}\BibitemShut {NoStop}%
\bibitem [{\citenamefont {Zhang}\ \emph {et~al.}(2015)\citenamefont {Zhang},
  \citenamefont {Qiao}, \citenamefont {Shi}, \citenamefont {Wu}, \citenamefont
  {Jiang},\ and\ \citenamefont {Tan}}]{Zhang2015}%
  \BibitemOpen
  \bibfield  {author} {\bibinfo {author} {\bibfnamefont {X.}~\bibnamefont
  {Zhang}}, \bibinfo {author} {\bibfnamefont {X.~F.}\ \bibnamefont {Qiao}},
  \bibinfo {author} {\bibfnamefont {W.}~\bibnamefont {Shi}}, \bibinfo {author}
  {\bibfnamefont {J.~B.}\ \bibnamefont {Wu}}, \bibinfo {author} {\bibfnamefont
  {D.~S.}\ \bibnamefont {Jiang}}, \ and\ \bibinfo {author} {\bibfnamefont
  {P.~H.}\ \bibnamefont {Tan}},\ }\href {\doibase 10.1039/C4CS00282B}
  {\bibfield  {journal} {\bibinfo  {journal} {Chem. Soc. Rev.}\ }\textbf
  {\bibinfo {volume} {44}},\ \bibinfo {pages} {2757} (\bibinfo {year}
  {2015})}\BibitemShut {NoStop}%
\bibitem [{\citenamefont {Lin}\ \emph {et~al.}(2016)\citenamefont {Lin},
  \citenamefont {Carvalho}, \citenamefont {Kahn}, \citenamefont {Lv},
  \citenamefont {Rao}, \citenamefont {Terrones}, \citenamefont {Pimenta},\ and\
  \citenamefont {Terrones}}]{Lin2016a}%
  \BibitemOpen
  \bibfield  {author} {\bibinfo {author} {\bibfnamefont {Z.}~\bibnamefont
  {Lin}}, \bibinfo {author} {\bibfnamefont {B.~R.}\ \bibnamefont {Carvalho}},
  \bibinfo {author} {\bibfnamefont {E.}~\bibnamefont {Kahn}}, \bibinfo {author}
  {\bibfnamefont {R.}~\bibnamefont {Lv}}, \bibinfo {author} {\bibfnamefont
  {R.}~\bibnamefont {Rao}}, \bibinfo {author} {\bibfnamefont {H.}~\bibnamefont
  {Terrones}}, \bibinfo {author} {\bibfnamefont {M.~A.}\ \bibnamefont
  {Pimenta}}, \ and\ \bibinfo {author} {\bibfnamefont {M.}~\bibnamefont
  {Terrones}},\ }\href@noop {} {\bibfield  {journal} {\bibinfo  {journal} {2D
  Materials}\ }\textbf {\bibinfo {volume} {3}} (\bibinfo {year}
  {2016})}\BibitemShut {NoStop}%
\bibitem [{\citenamefont {Song}\ \emph {et~al.}(2016)\citenamefont {Song},
  \citenamefont {Tan}, \citenamefont {Zhang}, \citenamefont {Wu}, \citenamefont
  {Sheng}, \citenamefont {Wan}, \citenamefont {Wang}, \citenamefont {Dai},\
  and\ \citenamefont {Tan}}]{Song2016}%
  \BibitemOpen
  \bibfield  {author} {\bibinfo {author} {\bibfnamefont {Q.~J.}\ \bibnamefont
  {Song}}, \bibinfo {author} {\bibfnamefont {Q.~H.}\ \bibnamefont {Tan}},
  \bibinfo {author} {\bibfnamefont {X.}~\bibnamefont {Zhang}}, \bibinfo
  {author} {\bibfnamefont {J.~B.}\ \bibnamefont {Wu}}, \bibinfo {author}
  {\bibfnamefont {B.~W.}\ \bibnamefont {Sheng}}, \bibinfo {author}
  {\bibfnamefont {Y.}~\bibnamefont {Wan}}, \bibinfo {author} {\bibfnamefont
  {X.~Q.}\ \bibnamefont {Wang}}, \bibinfo {author} {\bibfnamefont
  {L.}~\bibnamefont {Dai}}, \ and\ \bibinfo {author} {\bibfnamefont {P.~H.}\
  \bibnamefont {Tan}},\ }\href {\doibase 10.1103/PhysRevB.93.115409} {\bibfield
   {journal} {\bibinfo  {journal} {Physical Review B}\ }\textbf {\bibinfo
  {volume} {93}},\ \bibinfo {pages} {115409} (\bibinfo {year}
  {2016})}\BibitemShut {NoStop}%
\bibitem [{\citenamefont {Lee}\ \emph {et~al.}(2010)\citenamefont {Lee},
  \citenamefont {Yan}, \citenamefont {Brus}, \citenamefont {Heinz},
  \citenamefont {Hone},\ and\ \citenamefont {Ryu}}]{Lee2010}%
  \BibitemOpen
  \bibfield  {author} {\bibinfo {author} {\bibfnamefont {C.}~\bibnamefont
  {Lee}}, \bibinfo {author} {\bibfnamefont {H.}~\bibnamefont {Yan}}, \bibinfo
  {author} {\bibfnamefont {L.~E.}\ \bibnamefont {Brus}}, \bibinfo {author}
  {\bibfnamefont {T.~F.}\ \bibnamefont {Heinz}}, \bibinfo {author}
  {\bibfnamefont {J.}~\bibnamefont {Hone}}, \ and\ \bibinfo {author}
  {\bibfnamefont {S.}~\bibnamefont {Ryu}},\ }\href {\doibase 10.1021/nn1003937}
  {\bibfield  {journal} {\bibinfo  {journal} {ACS Nano}\ }\textbf {\bibinfo
  {volume} {4}},\ \bibinfo {pages} {2695} (\bibinfo {year} {2010})}\BibitemShut
  {NoStop}%
\bibitem [{\citenamefont {Li}\ \emph {et~al.}(2012)\citenamefont {Li},
  \citenamefont {Zhang}, \citenamefont {Yap}, \citenamefont {Tay},
  \citenamefont {Edwin}, \citenamefont {Olivier},\ and\ \citenamefont
  {Baillargeat}}]{Li2012}%
  \BibitemOpen
  \bibfield  {author} {\bibinfo {author} {\bibfnamefont {H.}~\bibnamefont
  {Li}}, \bibinfo {author} {\bibfnamefont {Q.}~\bibnamefont {Zhang}}, \bibinfo
  {author} {\bibfnamefont {C.~C.~R.}\ \bibnamefont {Yap}}, \bibinfo {author}
  {\bibfnamefont {B.~K.}\ \bibnamefont {Tay}}, \bibinfo {author} {\bibfnamefont
  {T.~H.~T.}\ \bibnamefont {Edwin}}, \bibinfo {author} {\bibfnamefont
  {A.}~\bibnamefont {Olivier}}, \ and\ \bibinfo {author} {\bibfnamefont
  {D.}~\bibnamefont {Baillargeat}},\ }\href {\doibase 10.1002/adfm.201102111}
  {\bibfield  {journal} {\bibinfo  {journal} {Advanced Functional Materials}\
  }\textbf {\bibinfo {volume} {22}},\ \bibinfo {pages} {1385} (\bibinfo {year}
  {2012})}\BibitemShut {NoStop}%
\bibitem [{\citenamefont {O'Brien}\ \emph {et~al.}(2015)\citenamefont
  {O'Brien}, \citenamefont {McEvoy}, \citenamefont {Hanlon}, \citenamefont
  {Hallam}, \citenamefont {Coleman},\ and\ \citenamefont
  {Duesberg}}]{OBrien2015}%
  \BibitemOpen
  \bibfield  {author} {\bibinfo {author} {\bibfnamefont {M.}~\bibnamefont
  {O'Brien}}, \bibinfo {author} {\bibfnamefont {N.}~\bibnamefont {McEvoy}},
  \bibinfo {author} {\bibfnamefont {D.}~\bibnamefont {Hanlon}}, \bibinfo
  {author} {\bibfnamefont {T.}~\bibnamefont {Hallam}}, \bibinfo {author}
  {\bibfnamefont {J.~N.}\ \bibnamefont {Coleman}}, \ and\ \bibinfo {author}
  {\bibfnamefont {G.~S.}\ \bibnamefont {Duesberg}},\ }\href@noop {} {\bibfield
  {journal} {\bibinfo  {journal} {Scientific Reports}\ }\textbf {\bibinfo
  {volume} {6}},\ \bibinfo {pages} {1} (\bibinfo {year} {2015})}\BibitemShut
  {NoStop}%
\bibitem [{\citenamefont {Zeng}\ \emph {et~al.}(2012)\citenamefont {Zeng},
  \citenamefont {Zhu}, \citenamefont {Liu}, \citenamefont {Fan}, \citenamefont
  {Cui},\ and\ \citenamefont {Zhang}}]{Zeng2012}%
  \BibitemOpen
  \bibfield  {author} {\bibinfo {author} {\bibfnamefont {H.}~\bibnamefont
  {Zeng}}, \bibinfo {author} {\bibfnamefont {B.}~\bibnamefont {Zhu}}, \bibinfo
  {author} {\bibfnamefont {K.}~\bibnamefont {Liu}}, \bibinfo {author}
  {\bibfnamefont {J.}~\bibnamefont {Fan}}, \bibinfo {author} {\bibfnamefont
  {X.}~\bibnamefont {Cui}}, \ and\ \bibinfo {author} {\bibfnamefont {Q.~M.}\
  \bibnamefont {Zhang}},\ }\href {\doibase 10.1103/PhysRevB.86.241301}
  {\bibfield  {journal} {\bibinfo  {journal} {Physical Review B}\ }\textbf
  {\bibinfo {volume} {86}},\ \bibinfo {pages} {1} (\bibinfo {year}
  {2012})}\BibitemShut {NoStop}%
\bibitem [{\citenamefont {Fox}\ \emph {et~al.}(2015)\citenamefont {Fox},
  \citenamefont {Zhou}, \citenamefont {Maguire}, \citenamefont {O'Neill},
  \citenamefont {O'Coile{\'{a}}in}, \citenamefont {Gatensby}, \citenamefont
  {Glushenkov}, \citenamefont {Tao}, \citenamefont {Duesberg}, \citenamefont
  {Shvets}, \citenamefont {Abid}, \citenamefont {Abid}, \citenamefont {Wu},
  \citenamefont {Chen}, \citenamefont {Coleman}, \citenamefont {Donegan},\ and\
  \citenamefont {Zhang}}]{Fox2015}%
  \BibitemOpen
  \bibfield  {author} {\bibinfo {author} {\bibfnamefont {D.~S.}\ \bibnamefont
  {Fox}}, \bibinfo {author} {\bibfnamefont {Y.}~\bibnamefont {Zhou}}, \bibinfo
  {author} {\bibfnamefont {P.}~\bibnamefont {Maguire}}, \bibinfo {author}
  {\bibfnamefont {A.}~\bibnamefont {O'Neill}}, \bibinfo {author} {\bibfnamefont
  {C.}~\bibnamefont {O'Coile{\'{a}}in}}, \bibinfo {author} {\bibfnamefont
  {R.}~\bibnamefont {Gatensby}}, \bibinfo {author} {\bibfnamefont {A.~M.}\
  \bibnamefont {Glushenkov}}, \bibinfo {author} {\bibfnamefont
  {T.}~\bibnamefont {Tao}}, \bibinfo {author} {\bibfnamefont {G.~S.}\
  \bibnamefont {Duesberg}}, \bibinfo {author} {\bibfnamefont {I.~V.}\
  \bibnamefont {Shvets}}, \bibinfo {author} {\bibfnamefont {M.}~\bibnamefont
  {Abid}}, \bibinfo {author} {\bibfnamefont {M.}~\bibnamefont {Abid}}, \bibinfo
  {author} {\bibfnamefont {H.~C.}\ \bibnamefont {Wu}}, \bibinfo {author}
  {\bibfnamefont {Y.}~\bibnamefont {Chen}}, \bibinfo {author} {\bibfnamefont
  {J.~N.}\ \bibnamefont {Coleman}}, \bibinfo {author} {\bibfnamefont {J.~F.}\
  \bibnamefont {Donegan}}, \ and\ \bibinfo {author} {\bibfnamefont
  {H.}~\bibnamefont {Zhang}},\ }\href {\doibase 10.1021/acs.nanolett.5b01673}
  {\bibfield  {journal} {\bibinfo  {journal} {Nano Letters}\ }\textbf {\bibinfo
  {volume} {15}},\ \bibinfo {pages} {5307} (\bibinfo {year}
  {2015})}\BibitemShut {NoStop}%
\bibitem [{\citenamefont {Ko}\ \emph {et~al.}(2016)\citenamefont {Ko},
  \citenamefont {Jeong}, \citenamefont {Kim}, \citenamefont {Lee},
  \citenamefont {Kim}, \citenamefont {Lee}, \citenamefont {Ryu}, \citenamefont
  {Park}, \citenamefont {Kim}, \citenamefont {Lee}, \citenamefont {Lee},
  \citenamefont {Lee},\ and\ \citenamefont {Ryu}}]{Ko2016}%
  \BibitemOpen
  \bibfield  {author} {\bibinfo {author} {\bibfnamefont {T.~Y.}\ \bibnamefont
  {Ko}}, \bibinfo {author} {\bibfnamefont {A.}~\bibnamefont {Jeong}}, \bibinfo
  {author} {\bibfnamefont {W.}~\bibnamefont {Kim}}, \bibinfo {author}
  {\bibfnamefont {J.}~\bibnamefont {Lee}}, \bibinfo {author} {\bibfnamefont
  {Y.}~\bibnamefont {Kim}}, \bibinfo {author} {\bibfnamefont {J.~E.}\
  \bibnamefont {Lee}}, \bibinfo {author} {\bibfnamefont {G.~H.}\ \bibnamefont
  {Ryu}}, \bibinfo {author} {\bibfnamefont {K.}~\bibnamefont {Park}}, \bibinfo
  {author} {\bibfnamefont {D.}~\bibnamefont {Kim}}, \bibinfo {author}
  {\bibfnamefont {Z.}~\bibnamefont {Lee}}, \bibinfo {author} {\bibfnamefont
  {M.~H.}\ \bibnamefont {Lee}}, \bibinfo {author} {\bibfnamefont
  {C.}~\bibnamefont {Lee}}, \ and\ \bibinfo {author} {\bibfnamefont
  {S.}~\bibnamefont {Ryu}},\ }\href {\doibase 10.1088/2053-1583/4/1/014003}
  {\bibfield  {journal} {\bibinfo  {journal} {2D Materials}\ }\textbf {\bibinfo
  {volume} {4}},\ \bibinfo {pages} {014003} (\bibinfo {year}
  {2016})}\BibitemShut {NoStop}%
\bibitem [{\citenamefont {Choudhary}\ \emph {et~al.}(2016)\citenamefont
  {Choudhary}, \citenamefont {Islam}, \citenamefont {Kang}, \citenamefont
  {Tetard}, \citenamefont {Jung},\ and\ \citenamefont
  {Khondaker}}]{Choudhary2016}%
  \BibitemOpen
  \bibfield  {author} {\bibinfo {author} {\bibfnamefont {N.}~\bibnamefont
  {Choudhary}}, \bibinfo {author} {\bibfnamefont {M.~R.}\ \bibnamefont
  {Islam}}, \bibinfo {author} {\bibfnamefont {N.}~\bibnamefont {Kang}},
  \bibinfo {author} {\bibfnamefont {L.}~\bibnamefont {Tetard}}, \bibinfo
  {author} {\bibfnamefont {Y.}~\bibnamefont {Jung}}, \ and\ \bibinfo {author}
  {\bibfnamefont {S.~I.}\ \bibnamefont {Khondaker}},\ }\href@noop {} {\bibfield
   {journal} {\bibinfo  {journal} {Journal of Physics Condensed Matter}\
  }\textbf {\bibinfo {volume} {28}} (\bibinfo {year} {2016})}\BibitemShut
  {NoStop}%
\bibitem [{\citenamefont {Maguire}\ \emph {et~al.}(2018)\citenamefont
  {Maguire}, \citenamefont {Fox}, \citenamefont {Zhou}, \citenamefont {Wang},
  \citenamefont {Brien}, \citenamefont {Jadwiszczak}, \citenamefont {Cullen},
  \citenamefont {McManus}, \citenamefont {Bateman}, \citenamefont {Mcevoy},
  \citenamefont {Duesberg},\ and\ \citenamefont {Zhang}}]{Maguire2018}%
  \BibitemOpen
  \bibfield  {author} {\bibinfo {author} {\bibfnamefont {P.}~\bibnamefont
  {Maguire}}, \bibinfo {author} {\bibfnamefont {D.~S.}\ \bibnamefont {Fox}},
  \bibinfo {author} {\bibfnamefont {Y.}~\bibnamefont {Zhou}}, \bibinfo {author}
  {\bibfnamefont {Q.}~\bibnamefont {Wang}}, \bibinfo {author} {\bibfnamefont
  {M.~O.}\ \bibnamefont {Brien}}, \bibinfo {author} {\bibfnamefont
  {J.}~\bibnamefont {Jadwiszczak}}, \bibinfo {author} {\bibfnamefont {C.~P.}\
  \bibnamefont {Cullen}}, \bibinfo {author} {\bibfnamefont {J.}~\bibnamefont
  {McManus}}, \bibinfo {author} {\bibfnamefont {S.}~\bibnamefont {Bateman}},
  \bibinfo {author} {\bibfnamefont {N.}~\bibnamefont {Mcevoy}}, \bibinfo
  {author} {\bibfnamefont {G.~S.}\ \bibnamefont {Duesberg}}, \ and\ \bibinfo
  {author} {\bibfnamefont {H.}~\bibnamefont {Zhang}},\ }\href@noop {}
  {\bibfield  {journal} {\bibinfo  {journal} {Physical Review B}\ }\textbf
  {\bibinfo {volume} {98}},\ \bibinfo {pages} {134109} (\bibinfo {year}
  {2018})}\BibitemShut {NoStop}%
\bibitem [{\citenamefont {Zhou}\ \emph {et~al.}(2016)\citenamefont {Zhou},
  \citenamefont {Maguire}, \citenamefont {Jadwiszczak}, \citenamefont
  {Muruganathan}, \citenamefont {Mizuta},\ and\ \citenamefont
  {Zhang}}]{Zhou2016a}%
  \BibitemOpen
  \bibfield  {author} {\bibinfo {author} {\bibfnamefont {Y.}~\bibnamefont
  {Zhou}}, \bibinfo {author} {\bibfnamefont {P.}~\bibnamefont {Maguire}},
  \bibinfo {author} {\bibfnamefont {J.}~\bibnamefont {Jadwiszczak}}, \bibinfo
  {author} {\bibfnamefont {M.}~\bibnamefont {Muruganathan}}, \bibinfo {author}
  {\bibfnamefont {H.}~\bibnamefont {Mizuta}}, \ and\ \bibinfo {author}
  {\bibfnamefont {H.}~\bibnamefont {Zhang}},\ }\href {\doibase
  10.1088/0957-4484/27/32/325302} {\bibfield  {journal} {\bibinfo  {journal}
  {Nanotechnology}\ }\textbf {\bibinfo {volume} {27}},\ \bibinfo {pages}
  {325302} (\bibinfo {year} {2016})}\BibitemShut {NoStop}%
\bibitem [{\citenamefont {Nanda}\ \emph {et~al.}(2015)\citenamefont {Nanda},
  \citenamefont {Goswami}, \citenamefont {Watanabe}, \citenamefont
  {Taniguchi},\ and\ \citenamefont {Alkemade}}]{Nanda2015}%
  \BibitemOpen
  \bibfield  {author} {\bibinfo {author} {\bibfnamefont {G.}~\bibnamefont
  {Nanda}}, \bibinfo {author} {\bibfnamefont {S.}~\bibnamefont {Goswami}},
  \bibinfo {author} {\bibfnamefont {K.}~\bibnamefont {Watanabe}}, \bibinfo
  {author} {\bibfnamefont {T.}~\bibnamefont {Taniguchi}}, \ and\ \bibinfo
  {author} {\bibfnamefont {P.~F.~A.}\ \bibnamefont {Alkemade}},\ }\href@noop {}
  {\bibfield  {journal} {\bibinfo  {journal} {Nano Letters}\ }\textbf {\bibinfo
  {volume} {15}},\ \bibinfo {pages} {4006} (\bibinfo {year}
  {2015})}\BibitemShut {NoStop}%
\bibitem [{\citenamefont {Iberi}\ \emph {et~al.}(2016)\citenamefont {Iberi},
  \citenamefont {Liangbo}, \citenamefont {Ievlev}, \citenamefont {Stanford},
  \citenamefont {Lin}, \citenamefont {Li}, \citenamefont {Mahjouri-Samani},
  \citenamefont {Jesse}, \citenamefont {Sumpter}, \citenamefont {Kalinin},
  \citenamefont {Joy}, \citenamefont {Xiao}, \citenamefont {Belianinov},\ and\
  \citenamefont {Ovchinnikova}}]{Iberi2016}%
  \BibitemOpen
  \bibfield  {author} {\bibinfo {author} {\bibfnamefont {V.}~\bibnamefont
  {Iberi}}, \bibinfo {author} {\bibfnamefont {L.}~\bibnamefont {Liangbo}},
  \bibinfo {author} {\bibfnamefont {A.~V.}\ \bibnamefont {Ievlev}}, \bibinfo
  {author} {\bibfnamefont {M.~G.}\ \bibnamefont {Stanford}}, \bibinfo {author}
  {\bibfnamefont {M.~W.}\ \bibnamefont {Lin}}, \bibinfo {author} {\bibfnamefont
  {X.}~\bibnamefont {Li}}, \bibinfo {author} {\bibfnamefont {M.}~\bibnamefont
  {Mahjouri-Samani}}, \bibinfo {author} {\bibfnamefont {S.}~\bibnamefont
  {Jesse}}, \bibinfo {author} {\bibfnamefont {B.~G.}\ \bibnamefont {Sumpter}},
  \bibinfo {author} {\bibfnamefont {S.~V.}\ \bibnamefont {Kalinin}}, \bibinfo
  {author} {\bibfnamefont {D.~C.}\ \bibnamefont {Joy}}, \bibinfo {author}
  {\bibfnamefont {K.}~\bibnamefont {Xiao}}, \bibinfo {author} {\bibfnamefont
  {A.}~\bibnamefont {Belianinov}}, \ and\ \bibinfo {author} {\bibfnamefont
  {O.~S.}\ \bibnamefont {Ovchinnikova}},\ }\href@noop {} {\bibfield  {journal}
  {\bibinfo  {journal} {Scientific Reports}\ }\textbf {\bibinfo {volume} {6}}
  (\bibinfo {year} {2016})}\BibitemShut {NoStop}%
\bibitem [{\citenamefont {Stanford}\ \emph {et~al.}(2016)\citenamefont
  {Stanford}, \citenamefont {Pudasaini}, \citenamefont {Belianinov},
  \citenamefont {Cross}, \citenamefont {Noh}, \citenamefont {Koehler},
  \citenamefont {Mandrus}, \citenamefont {Duscher}, \citenamefont {Rondinone},
  \citenamefont {Ivanov}, \citenamefont {Ward},\ and\ \citenamefont
  {Rack}}]{Stanford2016}%
  \BibitemOpen
  \bibfield  {author} {\bibinfo {author} {\bibfnamefont {M.~G.}\ \bibnamefont
  {Stanford}}, \bibinfo {author} {\bibfnamefont {P.~R.}\ \bibnamefont
  {Pudasaini}}, \bibinfo {author} {\bibfnamefont {A.}~\bibnamefont
  {Belianinov}}, \bibinfo {author} {\bibfnamefont {N.}~\bibnamefont {Cross}},
  \bibinfo {author} {\bibfnamefont {J.~H.}\ \bibnamefont {Noh}}, \bibinfo
  {author} {\bibfnamefont {M.~R.}\ \bibnamefont {Koehler}}, \bibinfo {author}
  {\bibfnamefont {D.~G.}\ \bibnamefont {Mandrus}}, \bibinfo {author}
  {\bibfnamefont {G.}~\bibnamefont {Duscher}}, \bibinfo {author} {\bibfnamefont
  {A.~J.}\ \bibnamefont {Rondinone}}, \bibinfo {author} {\bibfnamefont {I.~N.}\
  \bibnamefont {Ivanov}}, \bibinfo {author} {\bibfnamefont {T.~Z.}\
  \bibnamefont {Ward}}, \ and\ \bibinfo {author} {\bibfnamefont {P.~D.}\
  \bibnamefont {Rack}},\ }\href@noop {} {\bibfield  {journal} {\bibinfo
  {journal} {Scientific Reports}\ ,\ \bibinfo {pages} {27276}} (\bibinfo {year}
  {2016})}\BibitemShut {NoStop}%
\bibitem [{\citenamefont {Stanford}\ \emph {et~al.}(2017)\citenamefont
  {Stanford}, \citenamefont {Pudasaini}, \citenamefont {Gallmeier},
  \citenamefont {Cross}, \citenamefont {Liang}, \citenamefont {Oyedele},
  \citenamefont {Duscher}, \citenamefont {Mahjouri-Samani}, \citenamefont
  {Wang}, \citenamefont {Xiao}, \citenamefont {Geohegan}, \citenamefont
  {Belianinov}, \citenamefont {Sumpter},\ and\ \citenamefont
  {Rack}}]{Stanford2017b}%
  \BibitemOpen
  \bibfield  {author} {\bibinfo {author} {\bibfnamefont {M.~G.}\ \bibnamefont
  {Stanford}}, \bibinfo {author} {\bibfnamefont {P.~R.}\ \bibnamefont
  {Pudasaini}}, \bibinfo {author} {\bibfnamefont {E.~T.}\ \bibnamefont
  {Gallmeier}}, \bibinfo {author} {\bibfnamefont {N.}~\bibnamefont {Cross}},
  \bibinfo {author} {\bibfnamefont {L.}~\bibnamefont {Liang}}, \bibinfo
  {author} {\bibfnamefont {A.}~\bibnamefont {Oyedele}}, \bibinfo {author}
  {\bibfnamefont {G.}~\bibnamefont {Duscher}}, \bibinfo {author} {\bibfnamefont
  {M.}~\bibnamefont {Mahjouri-Samani}}, \bibinfo {author} {\bibfnamefont
  {K.}~\bibnamefont {Wang}}, \bibinfo {author} {\bibfnamefont {K.}~\bibnamefont
  {Xiao}}, \bibinfo {author} {\bibfnamefont {D.~B.}\ \bibnamefont {Geohegan}},
  \bibinfo {author} {\bibfnamefont {A.}~\bibnamefont {Belianinov}}, \bibinfo
  {author} {\bibfnamefont {B.~G.}\ \bibnamefont {Sumpter}}, \ and\ \bibinfo
  {author} {\bibfnamefont {P.~D.}\ \bibnamefont {Rack}},\ }\href@noop {}
  {\bibfield  {journal} {\bibinfo  {journal} {Advanced Functional Materials}\
  }\textbf {\bibinfo {volume} {27}} (\bibinfo {year} {2017})}\BibitemShut
  {NoStop}%
\bibitem [{\citenamefont {Nanda}\ \emph {et~al.}(2017)\citenamefont {Nanda},
  \citenamefont {Hlawacek}, \citenamefont {Goswami}, \citenamefont {Watanabe},
  \citenamefont {Taniguchi},\ and\ \citenamefont {Alkemade}}]{Nanda2017a}%
  \BibitemOpen
  \bibfield  {author} {\bibinfo {author} {\bibfnamefont {G.}~\bibnamefont
  {Nanda}}, \bibinfo {author} {\bibfnamefont {G.}~\bibnamefont {Hlawacek}},
  \bibinfo {author} {\bibfnamefont {S.}~\bibnamefont {Goswami}}, \bibinfo
  {author} {\bibfnamefont {K.}~\bibnamefont {Watanabe}}, \bibinfo {author}
  {\bibfnamefont {T.}~\bibnamefont {Taniguchi}}, \ and\ \bibinfo {author}
  {\bibfnamefont {P.~F.~A.}\ \bibnamefont {Alkemade}},\ }\href {\doibase
  10.1016/j.carbon.2017.04.062} {\bibfield  {journal} {\bibinfo  {journal}
  {Carbon}\ }\textbf {\bibinfo {volume} {119}},\ \bibinfo {pages} {419}
  (\bibinfo {year} {2017})}\BibitemShut {NoStop}%
\bibitem [{\citenamefont {O'Brien}\ \emph {et~al.}(2014)\citenamefont
  {O'Brien}, \citenamefont {McEvoy}, \citenamefont {Hallam}, \citenamefont
  {Kim}, \citenamefont {Berner}, \citenamefont {Hanlon}, \citenamefont {Lee},
  \citenamefont {Coleman},\ and\ \citenamefont {Duesberg}}]{OBrien2014}%
  \BibitemOpen
  \bibfield  {author} {\bibinfo {author} {\bibfnamefont {M.}~\bibnamefont
  {O'Brien}}, \bibinfo {author} {\bibfnamefont {N.}~\bibnamefont {McEvoy}},
  \bibinfo {author} {\bibfnamefont {T.}~\bibnamefont {Hallam}}, \bibinfo
  {author} {\bibfnamefont {H.~Y.}\ \bibnamefont {Kim}}, \bibinfo {author}
  {\bibfnamefont {N.~C.}\ \bibnamefont {Berner}}, \bibinfo {author}
  {\bibfnamefont {D.}~\bibnamefont {Hanlon}}, \bibinfo {author} {\bibfnamefont
  {K.}~\bibnamefont {Lee}}, \bibinfo {author} {\bibfnamefont {J.~N.}\
  \bibnamefont {Coleman}}, \ and\ \bibinfo {author} {\bibfnamefont {G.~S.}\
  \bibnamefont {Duesberg}},\ }\href@noop {} {\bibfield  {journal} {\bibinfo
  {journal} {Scientific Reports}\ }\textbf {\bibinfo {volume} {4}},\ \bibinfo
  {pages} {7374} (\bibinfo {year} {2014})}\BibitemShut {NoStop}%
\bibitem [{\citenamefont {Castellanos-Gomez}\ \emph {et~al.}(2012)\citenamefont
  {Castellanos-Gomez}, \citenamefont {Barkelid}, \citenamefont {Goossens},
  \citenamefont {Calado}, \citenamefont {{Van Der Zant}},\ and\ \citenamefont
  {Steele}}]{Gomez2012}%
  \BibitemOpen
  \bibfield  {author} {\bibinfo {author} {\bibfnamefont {A.}~\bibnamefont
  {Castellanos-Gomez}}, \bibinfo {author} {\bibfnamefont {M.}~\bibnamefont
  {Barkelid}}, \bibinfo {author} {\bibfnamefont {A.~M.}\ \bibnamefont
  {Goossens}}, \bibinfo {author} {\bibfnamefont {V.~E.}\ \bibnamefont
  {Calado}}, \bibinfo {author} {\bibfnamefont {H.~S.~J.}\ \bibnamefont {{Van
  Der Zant}}}, \ and\ \bibinfo {author} {\bibfnamefont {G.~A.}\ \bibnamefont
  {Steele}},\ }\href {\doibase 10.1021/nl301164v} {\bibfield  {journal}
  {\bibinfo  {journal} {Nano Letters}\ }\textbf {\bibinfo {volume} {12}},\
  \bibinfo {pages} {3187} (\bibinfo {year} {2012})}\BibitemShut {NoStop}%
\bibitem [{\citenamefont {Liang}\ \emph
  {et~al.}(2017{\natexlab{a}})\citenamefont {Liang}, \citenamefont {Zhang},
  \citenamefont {Sumpter}, \citenamefont {Tan}, \citenamefont {Tan},\ and\
  \citenamefont {Meunier}}]{Liang2017b}%
  \BibitemOpen
  \bibfield  {author} {\bibinfo {author} {\bibfnamefont {L.}~\bibnamefont
  {Liang}}, \bibinfo {author} {\bibfnamefont {J.}~\bibnamefont {Zhang}},
  \bibinfo {author} {\bibfnamefont {B.~G.}\ \bibnamefont {Sumpter}}, \bibinfo
  {author} {\bibfnamefont {Q.~H.}\ \bibnamefont {Tan}}, \bibinfo {author}
  {\bibfnamefont {P.~H.}\ \bibnamefont {Tan}}, \ and\ \bibinfo {author}
  {\bibfnamefont {V.}~\bibnamefont {Meunier}},\ }\href@noop {} {\bibfield
  {journal} {\bibinfo  {journal} {ACS Nano}\ }\textbf {\bibinfo {volume}
  {11}},\ \bibinfo {pages} {11777} (\bibinfo {year}
  {2017}{\natexlab{a}})}\BibitemShut {NoStop}%
\bibitem [{\citenamefont {Liu}\ \emph {et~al.}(2014)\citenamefont {Liu},
  \citenamefont {Zhang}, \citenamefont {Cao}, \citenamefont {Jin},
  \citenamefont {Qiu}, \citenamefont {Zhou}, \citenamefont {Zettle},
  \citenamefont {Yang}, \citenamefont {Louie},\ and\ \citenamefont
  {Wang}}]{Liu2014}%
  \BibitemOpen
  \bibfield  {author} {\bibinfo {author} {\bibfnamefont {K.}~\bibnamefont
  {Liu}}, \bibinfo {author} {\bibfnamefont {L.}~\bibnamefont {Zhang}}, \bibinfo
  {author} {\bibfnamefont {T.}~\bibnamefont {Cao}}, \bibinfo {author}
  {\bibfnamefont {C.}~\bibnamefont {Jin}}, \bibinfo {author} {\bibfnamefont
  {D.}~\bibnamefont {Qiu}}, \bibinfo {author} {\bibfnamefont {Q.}~\bibnamefont
  {Zhou}}, \bibinfo {author} {\bibfnamefont {A.}~\bibnamefont {Zettle}},
  \bibinfo {author} {\bibfnamefont {P.}~\bibnamefont {Yang}}, \bibinfo {author}
  {\bibfnamefont {S.~G.}\ \bibnamefont {Louie}}, \ and\ \bibinfo {author}
  {\bibfnamefont {F.}~\bibnamefont {Wang}},\ }\href@noop {} {\bibfield
  {journal} {\bibinfo  {journal} {Nature Communications}\ }\textbf {\bibinfo
  {volume} {5}} (\bibinfo {year} {2014})}\BibitemShut {NoStop}%
\bibitem [{\citenamefont {Mignuzzi}\ \emph {et~al.}(2015)\citenamefont
  {Mignuzzi}, \citenamefont {Pollard}, \citenamefont {Bonini}, \citenamefont
  {Brennan}, \citenamefont {Gilmore}, \citenamefont {Pimenta}, \citenamefont
  {Richards},\ and\ \citenamefont {Roy}}]{Mignuzzi2015}%
  \BibitemOpen
  \bibfield  {author} {\bibinfo {author} {\bibfnamefont {S.}~\bibnamefont
  {Mignuzzi}}, \bibinfo {author} {\bibfnamefont {A.~J.}\ \bibnamefont
  {Pollard}}, \bibinfo {author} {\bibfnamefont {N.}~\bibnamefont {Bonini}},
  \bibinfo {author} {\bibfnamefont {B.}~\bibnamefont {Brennan}}, \bibinfo
  {author} {\bibfnamefont {I.~S.}\ \bibnamefont {Gilmore}}, \bibinfo {author}
  {\bibfnamefont {M.~A.}\ \bibnamefont {Pimenta}}, \bibinfo {author}
  {\bibfnamefont {D.}~\bibnamefont {Richards}}, \ and\ \bibinfo {author}
  {\bibfnamefont {D.}~\bibnamefont {Roy}},\ }\href {\doibase
  10.1103/PhysRevB.91.195411} {\bibfield  {journal} {\bibinfo  {journal}
  {Physical Review B}\ }\textbf {\bibinfo {volume} {91}},\ \bibinfo {pages}
  {195411} (\bibinfo {year} {2015})}\BibitemShut {NoStop}%
\bibitem [{\citenamefont {Klein}\ \emph {et~al.}(2018)\citenamefont {Klein},
  \citenamefont {Kuc}, \citenamefont {Nolinder}, \citenamefont {Altzschner},
  \citenamefont {Wierzbowski}, \citenamefont {Sigger}, \citenamefont {Kreupl},
  \citenamefont {Finley}, \citenamefont {Wurstbauer}, \citenamefont
  {Holleitner},\ and\ \citenamefont {Kaniber}}]{Klein2017}%
  \BibitemOpen
  \bibfield  {author} {\bibinfo {author} {\bibfnamefont {J.}~\bibnamefont
  {Klein}}, \bibinfo {author} {\bibfnamefont {A.}~\bibnamefont {Kuc}}, \bibinfo
  {author} {\bibfnamefont {A.}~\bibnamefont {Nolinder}}, \bibinfo {author}
  {\bibfnamefont {M.}~\bibnamefont {Altzschner}}, \bibinfo {author}
  {\bibfnamefont {J.}~\bibnamefont {Wierzbowski}}, \bibinfo {author}
  {\bibfnamefont {F.}~\bibnamefont {Sigger}}, \bibinfo {author} {\bibfnamefont
  {F.}~\bibnamefont {Kreupl}}, \bibinfo {author} {\bibfnamefont {J.~J.}\
  \bibnamefont {Finley}}, \bibinfo {author} {\bibfnamefont {U.}~\bibnamefont
  {Wurstbauer}}, \bibinfo {author} {\bibfnamefont {A.~W.}\ \bibnamefont
  {Holleitner}}, \ and\ \bibinfo {author} {\bibfnamefont {M.}~\bibnamefont
  {Kaniber}},\ }\href@noop {} {\bibfield  {journal} {\bibinfo  {journal} {2D
  Materials}\ }\textbf {\bibinfo {volume} {5}},\ \bibinfo {pages} {011007}
  (\bibinfo {year} {2018})}\BibitemShut {NoStop}%
\bibitem [{\citenamefont {Splendiani}\ \emph {et~al.}(2010)\citenamefont
  {Splendiani}, \citenamefont {Sun}, \citenamefont {Zhang}, \citenamefont {Li},
  \citenamefont {Kim}, \citenamefont {Chim}, \citenamefont {Galli},\ and\
  \citenamefont {Wang}}]{Splendiani2010}%
  \BibitemOpen
  \bibfield  {author} {\bibinfo {author} {\bibfnamefont {A.}~\bibnamefont
  {Splendiani}}, \bibinfo {author} {\bibfnamefont {L.}~\bibnamefont {Sun}},
  \bibinfo {author} {\bibfnamefont {Y.}~\bibnamefont {Zhang}}, \bibinfo
  {author} {\bibfnamefont {T.}~\bibnamefont {Li}}, \bibinfo {author}
  {\bibfnamefont {J.}~\bibnamefont {Kim}}, \bibinfo {author} {\bibfnamefont
  {C.~Y.}\ \bibnamefont {Chim}}, \bibinfo {author} {\bibfnamefont
  {G.}~\bibnamefont {Galli}}, \ and\ \bibinfo {author} {\bibfnamefont
  {F.}~\bibnamefont {Wang}},\ }\href {\doibase 10.1021/nl903868w} {\bibfield
  {journal} {\bibinfo  {journal} {Nano Letters}\ }\textbf {\bibinfo {volume}
  {10}},\ \bibinfo {pages} {1271} (\bibinfo {year} {2010})}\BibitemShut
  {NoStop}%
\bibitem [{\citenamefont {Scheuschner}\ \emph {et~al.}(2014)\citenamefont
  {Scheuschner}, \citenamefont {Ochedowski}, \citenamefont {Kaulitz},
  \citenamefont {Gillen}, \citenamefont {Schleberger},\ and\ \citenamefont
  {Maultzsch}}]{Scheuschner2014}%
  \BibitemOpen
  \bibfield  {author} {\bibinfo {author} {\bibfnamefont {N.}~\bibnamefont
  {Scheuschner}}, \bibinfo {author} {\bibfnamefont {O.}~\bibnamefont
  {Ochedowski}}, \bibinfo {author} {\bibfnamefont {A.~M.}\ \bibnamefont
  {Kaulitz}}, \bibinfo {author} {\bibfnamefont {R.}~\bibnamefont {Gillen}},
  \bibinfo {author} {\bibfnamefont {M.}~\bibnamefont {Schleberger}}, \ and\
  \bibinfo {author} {\bibfnamefont {J.}~\bibnamefont {Maultzsch}},\ }\href
  {\doibase 10.1103/PhysRevB.89.125406} {\bibfield  {journal} {\bibinfo
  {journal} {Physical Review B}\ }\textbf {\bibinfo {volume} {89}},\ \bibinfo
  {pages} {125406} (\bibinfo {year} {2014})}\BibitemShut {NoStop}%
\bibitem [{\citenamefont {Tan}\ \emph {et~al.}(2012)\citenamefont {Tan},
  \citenamefont {Han}, \citenamefont {Zhao}, \citenamefont {Wu}, \citenamefont
  {Chang}, \citenamefont {Wang}, \citenamefont {Wang}, \citenamefont {Bonini},
  \citenamefont {Marzari}, \citenamefont {Pugno}, \citenamefont {Savini},
  \citenamefont {Lombardo},\ and\ \citenamefont {Ferrari}}]{Tan2012}%
  \BibitemOpen
  \bibfield  {author} {\bibinfo {author} {\bibfnamefont {P.~H.}\ \bibnamefont
  {Tan}}, \bibinfo {author} {\bibfnamefont {W.~P.}\ \bibnamefont {Han}},
  \bibinfo {author} {\bibfnamefont {W.~J.}\ \bibnamefont {Zhao}}, \bibinfo
  {author} {\bibfnamefont {Z.~H.}\ \bibnamefont {Wu}}, \bibinfo {author}
  {\bibfnamefont {K.}~\bibnamefont {Chang}}, \bibinfo {author} {\bibfnamefont
  {H.}~\bibnamefont {Wang}}, \bibinfo {author} {\bibfnamefont {Y.~F.}\
  \bibnamefont {Wang}}, \bibinfo {author} {\bibfnamefont {N.}~\bibnamefont
  {Bonini}}, \bibinfo {author} {\bibfnamefont {N.}~\bibnamefont {Marzari}},
  \bibinfo {author} {\bibfnamefont {N.}~\bibnamefont {Pugno}}, \bibinfo
  {author} {\bibfnamefont {G.}~\bibnamefont {Savini}}, \bibinfo {author}
  {\bibfnamefont {A.}~\bibnamefont {Lombardo}}, \ and\ \bibinfo {author}
  {\bibfnamefont {A.~C.}\ \bibnamefont {Ferrari}},\ }\href {\doibase
  10.1038/nmat3245} {\bibfield  {journal} {\bibinfo  {journal} {Nature
  Materials}\ }\textbf {\bibinfo {volume} {11}},\ \bibinfo {pages} {294}
  (\bibinfo {year} {2012})}\BibitemShut {NoStop}%
\bibitem [{\citenamefont {Kretschmer}\ \emph {et~al.}(2018)\citenamefont
  {Kretschmer}, \citenamefont {Maslov}, \citenamefont {Ghaderzadeh},
  \citenamefont {Ghorbani-Asl}, \citenamefont {Hlawacek},\ and\ \citenamefont
  {Krasheninnikov}}]{Kretschmer2018}%
  \BibitemOpen
  \bibfield  {author} {\bibinfo {author} {\bibfnamefont {S.}~\bibnamefont
  {Kretschmer}}, \bibinfo {author} {\bibfnamefont {M.}~\bibnamefont {Maslov}},
  \bibinfo {author} {\bibfnamefont {S.}~\bibnamefont {Ghaderzadeh}}, \bibinfo
  {author} {\bibfnamefont {M.}~\bibnamefont {Ghorbani-Asl}}, \bibinfo {author}
  {\bibfnamefont {G.}~\bibnamefont {Hlawacek}}, \ and\ \bibinfo {author}
  {\bibfnamefont {A.~V.}\ \bibnamefont {Krasheninnikov}},\ }\href@noop {}
  {\bibfield  {journal} {\bibinfo  {journal} {ACS Applied Materials \&
  Interfaces}\ } (\bibinfo {year} {2018})}\BibitemShut {NoStop}%
\bibitem [{\citenamefont {Parkin}\ \emph {et~al.}(2016)\citenamefont {Parkin},
  \citenamefont {Balan}, \citenamefont {Liang}, \citenamefont {Das},
  \citenamefont {Lamparski}, \citenamefont {Naylor}, \citenamefont
  {Rodr{\'{i}}guez-Manzo}, \citenamefont {Johnson}, \citenamefont {Meunier},\
  and\ \citenamefont {Drndi{\'{c}}}}]{Parkin2016}%
  \BibitemOpen
  \bibfield  {author} {\bibinfo {author} {\bibfnamefont {W.~M.}\ \bibnamefont
  {Parkin}}, \bibinfo {author} {\bibfnamefont {A.}~\bibnamefont {Balan}},
  \bibinfo {author} {\bibfnamefont {L.}~\bibnamefont {Liang}}, \bibinfo
  {author} {\bibfnamefont {P.~M.}\ \bibnamefont {Das}}, \bibinfo {author}
  {\bibfnamefont {M.}~\bibnamefont {Lamparski}}, \bibinfo {author}
  {\bibfnamefont {C.~H.}\ \bibnamefont {Naylor}}, \bibinfo {author}
  {\bibfnamefont {J.~A.}\ \bibnamefont {Rodr{\'{i}}guez-Manzo}}, \bibinfo
  {author} {\bibfnamefont {A.~T.~C.}\ \bibnamefont {Johnson}}, \bibinfo
  {author} {\bibfnamefont {V.}~\bibnamefont {Meunier}}, \ and\ \bibinfo
  {author} {\bibfnamefont {M.}~\bibnamefont {Drndi{\'{c}}}},\ }\href {\doibase
  10.1021/acsnano.5b07388} {\bibfield  {journal} {\bibinfo  {journal} {ACS
  Nano}\ }\textbf {\bibinfo {volume} {10}},\ \bibinfo {pages} {4134} (\bibinfo
  {year} {2016})}\BibitemShut {NoStop}%
\bibitem [{\citenamefont {Huang}\ \emph {et~al.}(2016)\citenamefont {Huang},
  \citenamefont {Liang}, \citenamefont {Ling}, \citenamefont {Puretzky},
  \citenamefont {Geohegan}, \citenamefont {Sumpter}, \citenamefont {Kong},
  \citenamefont {Meunier},\ and\ \citenamefont {Dresselhaus}}]{Huang2016}%
  \BibitemOpen
  \bibfield  {author} {\bibinfo {author} {\bibfnamefont {S.}~\bibnamefont
  {Huang}}, \bibinfo {author} {\bibfnamefont {L.}~\bibnamefont {Liang}},
  \bibinfo {author} {\bibfnamefont {X.}~\bibnamefont {Ling}}, \bibinfo {author}
  {\bibfnamefont {A.~A.}\ \bibnamefont {Puretzky}}, \bibinfo {author}
  {\bibfnamefont {D.~B.}\ \bibnamefont {Geohegan}}, \bibinfo {author}
  {\bibfnamefont {B.~G.}\ \bibnamefont {Sumpter}}, \bibinfo {author}
  {\bibfnamefont {J.}~\bibnamefont {Kong}}, \bibinfo {author} {\bibfnamefont
  {V.}~\bibnamefont {Meunier}}, \ and\ \bibinfo {author} {\bibfnamefont
  {M.~S.}\ \bibnamefont {Dresselhaus}},\ }\href {\doibase
  10.1021/acs.nanolett.5b05015} {\bibfield  {journal} {\bibinfo  {journal}
  {Nano Letters}\ }\textbf {\bibinfo {volume} {16}},\ \bibinfo {pages} {1435}
  (\bibinfo {year} {2016})}\BibitemShut {NoStop}%
\bibitem [{\citenamefont {Liang}\ \emph
  {et~al.}(2017{\natexlab{b}})\citenamefont {Liang}, \citenamefont {Puretzky},
  \citenamefont {Sumpter},\ and\ \citenamefont {Meunier}}]{Liang2017a}%
  \BibitemOpen
  \bibfield  {author} {\bibinfo {author} {\bibfnamefont {L.}~\bibnamefont
  {Liang}}, \bibinfo {author} {\bibfnamefont {A.~A.}\ \bibnamefont {Puretzky}},
  \bibinfo {author} {\bibfnamefont {B.~G.}\ \bibnamefont {Sumpter}}, \ and\
  \bibinfo {author} {\bibfnamefont {V.}~\bibnamefont {Meunier}},\ }\href@noop
  {} {\bibfield  {journal} {\bibinfo  {journal} {Nanoscale}\ }\textbf {\bibinfo
  {volume} {9}},\ \bibinfo {pages} {15340} (\bibinfo {year}
  {2017}{\natexlab{b}})}\BibitemShut {NoStop}%
\bibitem [{\citenamefont {Lahouij}\ \emph {et~al.}(2012)\citenamefont
  {Lahouij}, \citenamefont {Vacher}, \citenamefont {Martin},\ and\
  \citenamefont {Dassenoy}}]{Lahouij2012}%
  \BibitemOpen
  \bibfield  {author} {\bibinfo {author} {\bibfnamefont {I.}~\bibnamefont
  {Lahouij}}, \bibinfo {author} {\bibfnamefont {B.}~\bibnamefont {Vacher}},
  \bibinfo {author} {\bibfnamefont {J.~M.}\ \bibnamefont {Martin}}, \ and\
  \bibinfo {author} {\bibfnamefont {F.}~\bibnamefont {Dassenoy}},\ }\href@noop
  {} {\bibfield  {journal} {\bibinfo  {journal} {Wear}\ }\textbf {\bibinfo
  {volume} {296}},\ \bibinfo {pages} {558} (\bibinfo {year}
  {2012})}\BibitemShut {NoStop}%
\bibitem [{\citenamefont {Rueden}\ \emph {et~al.}(2017)\citenamefont {Rueden},
  \citenamefont {Schindelin}, \citenamefont {Hiner}, \citenamefont {DeZonia},
  \citenamefont {Walter}, \citenamefont {Arena},\ and\ \citenamefont
  {Eliceiri}}]{Rueden2017}%
  \BibitemOpen
  \bibfield  {author} {\bibinfo {author} {\bibfnamefont {C.}~\bibnamefont
  {Rueden}}, \bibinfo {author} {\bibfnamefont {J.}~\bibnamefont {Schindelin}},
  \bibinfo {author} {\bibfnamefont {M.}~\bibnamefont {Hiner}}, \bibinfo
  {author} {\bibfnamefont {B.}~\bibnamefont {DeZonia}}, \bibinfo {author}
  {\bibfnamefont {A.}~\bibnamefont {Walter}}, \bibinfo {author} {\bibfnamefont
  {E.}~\bibnamefont {Arena}}, \ and\ \bibinfo {author} {\bibfnamefont
  {K.}~\bibnamefont {Eliceiri}},\ }\href@noop {} {\bibfield  {journal}
  {\bibinfo  {journal} {BMC Bioinformatics}\ }\textbf {\bibinfo {volume}
  {18}},\ \bibinfo {pages} {1} (\bibinfo {year} {2017})}\BibitemShut {NoStop}%
\end{thebibliography}%

\end{document}